\documentclass[journal]{IEEEtran}
\usepackage[dvipsnames]{xcolor}
\usepackage{amsmath,amsfonts,graphicx,float}
\usepackage{xcolor}
\usepackage{multicol}
\usepackage{array}
\usepackage{booktabs}
\usepackage{ragged2e}
\usepackage{multicol}
\usepackage{multirow,bigstrut}
\usepackage{color}
\usepackage{tikz}
\usetikzlibrary{shapes,arrows}
\usepackage{xspace}
\usepackage{graphicx}
\usepackage[justification=centering]{caption}
\usepackage{subcaption}
\usepackage{lipsum}
\usepackage{cite}
\usepackage[colorlinks = fasle, linkcolor = blue, urlcolor  = blue, citecolor = green, anchorcolor = blue]{hyperref}
\usepackage{algorithm2e}
\usetikzlibrary{shapes,fit}
\usepackage{multirow}
\usepackage[pscoord]{eso-pic}
\usetikzlibrary{intersections}
\usepackage{cleveref}
\usepackage{comment}

\newcolumntype{L}{>{\centering\arraybackslash}m{0.035\textwidth}}
\newcolumntype{Q}{>{\centering\arraybackslash}m{0.048\textwidth}}
\newcolumntype{M}{>{\centering\arraybackslash}m{0.100\textwidth}}
\newcolumntype{N}{>{\centering\arraybackslash}m{0.050\textwidth}}
\newcolumntype{S}{>{\centering\arraybackslash}m{0.075\textwidth}}
\newcolumntype{K}{>{\centering\arraybackslash}m{0.15\textwidth}}
\newcolumntype{O}{>{\centering\arraybackslash}m{0.065\textwidth}}
\newcolumntype{J}{>{\centering\arraybackslash}m{0.2\textwidth}}
\newcolumntype{F}{>{\centering\arraybackslash}m{0.045\textwidth}}

\begin{document}
\title{Subjective and Objective Quality Assessment Methods of Stereoscopic Videos with Visibility Affecting Distortions}

\author{Sria~Biswas$^1$,~Balasubramanyam~Appina$^2$,~\textit{Member}~\textit{IEEE},~Priyanka~Kokil$^1$,~\textit{Senior~Member}~\textit{IEEE},\\~Sumohana~S~Channappayya$^3$,~\textit{Senior~Member}~\textit{IEEE}.

\thanks{The research reported in this paper was supported in part by the Department of Science and Technology (DST) -- Science and Engineering Research Board, Government of India under Grant SRG/2020/000336. The work was also supported by the DST under the National Mission on Interdisciplinary Cyber-Physical Systems (NM-ICPS), Govt. of India, Technology Innovation Hub (TiHAN), IIT Hyderabad, under the grant TiHAN-IITH/03/2021-22/46(4).\\
$^1$ The authors are with the Department of Electronics and Communication Engineering, Indian Institute of Information Technology, Design and Manufacturing Kancheepuram, Chennai, Tamil Nadu-600127, India. e-mail: \{ec21d0002,~priyanka\}@iiitdm.ac.in.

$^2$ The author is with the Department of Electrical Engineering, Indian Institute of Technology Indore, 453552, India. e-mail: appina@iiti.ac.in.

$^3$ The author is with the Lab for Video and Image Analysis (LFOVIA), Department of Electrical Engineering, Indian Institute of Technology Hyderabad, Kandi,
India, 502285. e-mail: sumohana@iith.ac.in.}}
\date{\today}
\maketitle
\begin{abstract}
We present two major contributions in this work: 1) we create a full HD resolution stereoscopic (S3D) video dataset comprised of 12 reference and 360 distorted videos. The test stimuli are produced by simulating the five levels of fog and haze ambiances on the pristine left and right video sequences. We perform subjective analysis on the created video dataset with 24 viewers and compute Difference Mean Opinion Scores (DMOS) as quality representative of the dataset, 2) an Opinion Unaware (OU) and Distortion Unaware (DU) video quality assessment model is developed for S3D videos. We construct cyclopean frames from the individual views of an S3D video and partition them into nonoverlapping blocks. We analyze the Natural Scene Statistics (NSS) of all patches of pristine and test videos, and empirically model the NSS features with Univariate Generalized Gaussian Distribution (UGGD). We compute UGGD model parameters ($\alpha, \beta$) at multiple spatial scales and multiple orientations of spherical steerable pyramid decomposition and show that the UGGD parameters are distortion discriminable. Further, we perform Multivariate Gaussian (MVG) modeling on the pristine and distorted video feature sets and compute the corresponding mean vectors and covariance matrices of MVG fits. We compute the Bhattacharyya distance measure between mean vectors and covariance matrices to estimate the perceptual deviation of a test video from pristine video set. Finally, we pool both distance measures to estimate the overall quality score of an S3D video. The performance of the proposed objective algorithm is verified on the popular S3D video datasets such as IRCCYN, LFOVIAS3DPh1, LFOVIAS3DPh2 and the proposed VAD stereo dataset. The algorithm delivers consistent performance across all datasets and shows competitive performance against off-the-shelf 2D and 3D image and video quality assessment algorithms.
\end{abstract}
\section{Introduction}
\label{sec:intro}
The rapid upgrade in human lifestyle accompanied by the growing population of the world has resulted in an increasing number of vehicles on our streets. This has also led to a steep escalation in the number of vehicular collisions and accidents due to various reasons. Every year, about 1.3 million people die in traffic accidents and 20-50 million others are injured or disabled~\cite{website:who_accident}. The major reason behind such accidents are environmental factors and weather conditions like fog, haze, rain, snow, smoke, etc., which lead to poor visibility circumstances, are completely unavoidable and totally out of human control~\cite{website:fog_accident}. To ensure safe driving practices, there is a huge demand for the development of technologies to aid Advanced Driver Assistance Systems (ADAS). The main aim of these technologies is to successfully mimic the view that is perceived by the Human Visual System (HVS) and overcome the complications occurring due to human error. Our HVS considers left and right scenes as inputs and fuses both scenes to create an illusion of a stereoscopic (S3D or 3D) scene. An S3D scene consists of depth information along with spatial and temporal information which leads to a better Quality of Experience (QoE). This helps us to develop more safety-driven decision-making technologies for autonomous driving systems.

However, the creation of any 3D content comes with a few vital constraints~\cite{lambooij2011evaluation}. In order to produce 3D multimedia, we need high spatial resolution cameras, huge volumes of storage space and high data rates, and specially designed 3D displays are required for viewing purposes. The encoding/decoding process results in different types of errors in the source generated content. All of these factors contribute to a substantial degradation in the perceptual QoE of a viewer, and create a necessary requirement of quality assessment (QA) models. 

Quality assessment (QA) models are classified into subjective and objective assessment methods. The subjective assessment procedure involves human observers to assess the perceptual quality, which is a time-consuming and cumbersome method. Despite their hectic nature, subjective evaluation plays a vital role since the video content is created for human consumption, and the human opinion scores are crucial benchmarks for objective evaluation algorithms. The objective quality assessment method performs the automatic quality prediction of a video that mimics the human assessment scores. These methods are classified into three categories based on the utilization of pristine information content. In full reference (FR) models, the pristine information is fully utilized. Reduced reference (RR) models require partial information of pristine content, while no reference (NR) models do not require any pristine information in estimating the quality of a test video. 
\begin{figure*}[!htbp]
\captionsetup[subfigure]{justification=centering}
\centering
\begin{subfigure}[b]{0.2\textwidth}
\includegraphics[width=3.3cm]{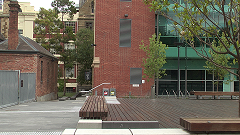} 
\subcaption{\small RMIT University Courtyard.}
\label{fig:4lref}
\end{subfigure}
\begin{subfigure}[b]{0.2\textwidth}
\includegraphics[width=3.3cm]{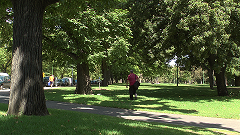}
\subcaption{\small Domain Parklands.\newline}
\label{fig:5lref}
\end{subfigure}
\begin{subfigure}[b]{0.2\textwidth}
\includegraphics[width=3.3cm]{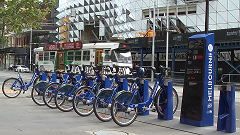} 
\subcaption{\small Melbourne Bicycle Stand (Swanston Street).}
\label{fig:7lref}
\end{subfigure}
\begin{subfigure}[b]{0.2\textwidth}
\includegraphics[width=3.3cm]{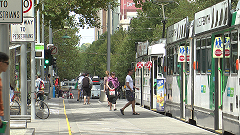}
\subcaption{\small Swanston Street Tram Stop.}
\label{fig:8lref}
\end{subfigure}
\\
\begin{subfigure}[b]{0.2\textwidth}
\includegraphics[width=3.3cm]{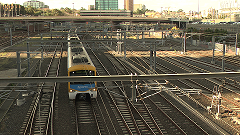}
\subcaption{\small Flinders Street Station.}
\label{fig:9lref}
\end{subfigure}
\begin{subfigure}[b]{0.2\textwidth}
\includegraphics[width=3.3cm]{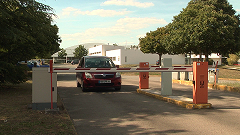}
\subcaption{\small NAMA3DS1 Barrier.}
\label{fig:10lref}
\end{subfigure}
\begin{subfigure}[b]{0.2\textwidth}
\includegraphics[width=3.3cm]{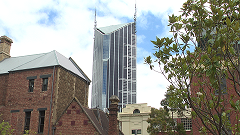}
\subcaption{\small Melbourne Tower.}
\label{fig:12lref}
\end{subfigure}
\begin{subfigure}[b]{0.2\textwidth}
\includegraphics[width=3.3cm]{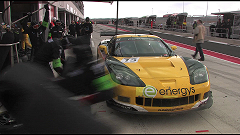}
\subcaption{\small Pit Stop Car Race.\newline}
\label{fig:13lref}
\end{subfigure}
\\
\begin{subfigure}[b]{0.2\textwidth}
\includegraphics[width=3.3cm]{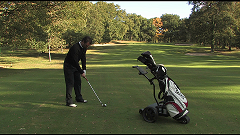}
\subcaption{\small Golf.}
\label{fig:14lref}
\end{subfigure}
\begin{subfigure}[b]{0.2\textwidth}
\includegraphics[width=3.3cm]{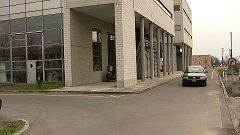} 
\subcaption{\small Car.}
\label{fig:6lref}
\end{subfigure}
\begin{subfigure}[b]{0.2\textwidth}
\includegraphics[width=3.3cm]{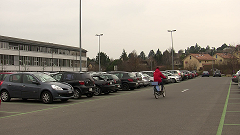}
\subcaption{\small Bike.}
\label{fig:15lref}
\end{subfigure}
\begin{subfigure}[b]{0.2\textwidth}
\includegraphics[width=3.3cm]{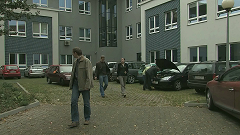}
\subcaption{\small Campus.}
\label{fig:11lref}
\end{subfigure}
\caption{Illustration of the $100^{th}$ frame from the left view of pristine S3D video.}
\label{fig:all reference video frames}
\end{figure*}

This article covers both subjective and objective methods for S3D videos. In subjective study, we create an S3D video dataset with 12 reference and 360 distorted videos, and the test stimuli are generated by producing the Visibility Affecting Distortions (VAD) such as fog and haze ambiances on the left and right views of an S3D video. We named our dataset as VAD stereo dataset. In objective assessment, we propose an OU and DU (i.e. completely blind) NR QA model for S3D videos based on analyzing the NSS features of generated cyclopean frames of an S3D video. We call our algorithm as $\text{C}$ompletely $\text{B}$lind $\text{S}$tereoscopic Video Quality $\text{E}$valuator (CBSE).

The paper is organized as follows. Section~\ref{sec:background} presents current literature on subjective and objective methods of S3D videos. Section~\ref{sec:subjectivequalityasessment} explains the subjective assessment on proposed VAD dataset. Section~\ref{sec:objectivequalityasessment} describes the proposed CBSE model. Section~\ref{sec:ResultsandDiscussion} provides the authenticity of the proposed dataset and discusses CBSE model performance. Section~\ref{sec: conclusion} summarizes our entire work, provides valid remarks and explores the scope of future improvements.
\section{Related Works}
\label{sec:background}
In this section, we present the latest literature on subjective quality assessment of VAD artefacts based 2D datasets and stereoscopic 3D datasets, and objective quality assessment methods for S3D videos.
\subsection{Subjective Quality Assessment}
\label{sec: subjective}
Research communities have an impressive collection of 2D image/video quality assessment (IQA/VQA) datasets when compared to S3D IQA/VQA datasets~\cite{Winkler2001,rohaly2000video}. The creation of 3D content has been slow due to the requirement of highly expensive professional stereo cameras, scene capturing constraints and complexities, storage issues, etc. The same applies to VAD artefacts also. 
\subsubsection{VAD artefacts based 2D image and video datasets}
Several works~\cite{li2017haze,juneja2021systematic} present a review on VAD artefacts based 2D image and video datasets which explore the properties, key ideas and specifications of the datasets. Also, these works emphasize the contributions and gaps in publicly available datasets. Narasimhan~\textit{et al.}~\cite{narasimhan2002all} were the first research group to propose a 2D fog image dataset. They have varied the scattering parameter value and chrominance decomposition to produce different foggy distorted images. Later, Tarel~\textit{et al.} designed multiple foggy image datasets~\cite{jpt-iv10,jpt-itsm12} specially for the ADAS based vehicles. The pristine images were created using the Sivic Software. Khoury~\textit{et al.}~\cite{el2016color} created a hazy 2D image dataset with a diverse combination of indoor scenes, and the haze ambiance was simulated using FOGBURST 1500 fog machine. The Haze Realistic Dataset (HazeRD) introduced by Zhang~\textit{et al.}~\cite{zhang2017hazerd} contained artificially simulated fog and haze images. Ancuti~\textit{et al.}~\cite{ancuti2018hazeO,ancuti2018hazeI} created indoor and outdoor scene hazy datasets based on simulating the hazy environment by using two LSM1500 PRO 1500 W haze machines.

Apart from images, a few VAD artefacts based 2D video datasets~\cite{pal2018visibility, 8451572, 9082053} are also publicly available. Ren~\textit{et al.}~\cite{ren2018deep} synthesized a hazy dataset by randomly selecting 100 videos from the NYU depth dataset~\cite{silberman2012indoor}. Zhang~\textit{et al.}~\cite{zhang2021learning} created the REal-world VIdeo DEhazing (REVIDE) dataset which comprises of various hazy and haze-free video sequences. 
\subsubsection{Stereoscopic video datasets}
We present a comprehensive review on S3D video datasets. Aflaki~\textit{et al.}~\cite{aflaki2010subjective} created an S3D video dataset to study the effects of asymmetric encoding on perceptual quality. They concluded that asymmetric encoding provides better bitrate savings when compared to symmetric encoding. De Silva~\textit{et al.}~\cite{de2013toward} proposed an S3D video dataset with 14 pristine and 116 distorted videos. The test stimuli are a combination of H.264 and H.265 compression artefacts. They concluded that higher quantization step sizes resulted in greater degradation in perceptual quality than lower quantization step sizes. Hewage~\textit{et al.}~\cite{Hewage2013} created an S3D video dataset with a combination of 9 pristine and 54 distorted videos. The packet loss artefacts were simulated at random locations using the JM reference software. They concluded that the overall perceptual quality of an S3D video is a function of both left and right video qualities. 

Several works~\cite{chen2012,appina2017subjective,wang2011subjective,Urvoy2012,appina2019subjective} proposed S3D video datasets by adding spatiotemporal distortions and depth errors. They concluded that human assessment scores are correlated with spatial quality but have a different relationship with disparity quality. Also, they stated that compression artefacts severely affect the S3D video quality which are having smaller depth ranges compared to the large depth ranges. Cheng~\textit{et al.}~\cite{cheng2012rmit3dv} created an S3D video dataset to analyze the movie postprocessing artefacts. Ha and Kim~\textit{et al.}~\cite{Ha2011} created an S3D video dataset to analyze the human perceptual affecting parameters such as visual quality, depth perception and visual discomfort. Chen~\textit{et al.}~\cite{chen2018blind} performed a subjective study on video compression artefacts to examine the relationship between S3D video quality, QoE and depth quality. Dehkordi~\textit{et al.}~\cite{Dehkordi2014} created an S3D video dataset and performed the subjective assessment to investigate the binocular rivalry and perceptual view dominance in an S3D scene. They concluded that binocular dominance seems to prevail during perception of an 3D scene when high quality views are compared with low quality views.
\begin{figure}[!htbp]
\centering
\begin{subfigure}[b]{0.23\textwidth}
\includegraphics[height=3cm,width=4cm]{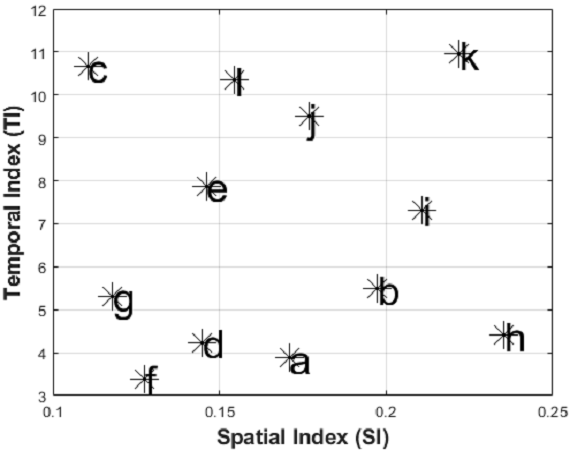}
\subcaption{\small Spatial and temporal indices (SI and TI) plot.}
\label{fig:SITI}
\end{subfigure}
\begin{subfigure}[b]{0.23\textwidth}
\includegraphics[height=3cm,width=4cm]{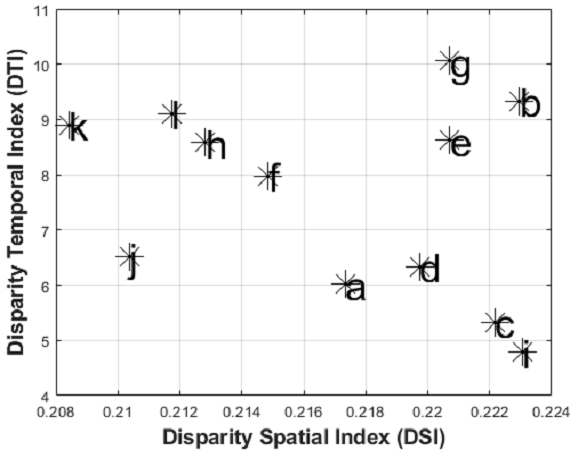}
\subcaption{\small Disparity SI and TI (DSI and DTI) plot.}
\label{fig:DSIDTI}
\end{subfigure}
\caption{The variation of Spatial Index (SI) and Temporal Index (TI) scores of reference videos, and disparity SI and TI (DSI and DTI) scores of the corresponding pristine S3D videos.}
\label{fig:sitiMain}
\end{figure}

The VAD artefacts based 2D image and video datasets~\cite{li2017haze, juneja2021systematic, narasimhan2002all, jpt-iv10, jpt-itsm12, el2016color, zhang2017hazerd, ancuti2018hazeO, ancuti2018hazeI, pal2018visibility, 8451572, 9082053, ren2018deep, silberman2012indoor, zhang2021learning} are produced by performing the artificial simulation of perceptual ambiance levels on image and video. The S3D video datasets~\cite{aflaki2010subjective, de2013toward, Hewage2013, chen2012, appina2017subjective, wang2011subjective, Urvoy2012, appina2019subjective, cheng2012rmit3dv, Ha2011, chen2018blind, Dehkordi2014} are proposed based on including the different types of spatiotemporal and depth distortions and visual discomfort artefacts. These works show a significant contribution in understanding the effect of different distortions and artefacts on the perceptual quality of a given scene content. However, none of these methods specifically consider the creation of VAD artefacts based S3D video content and explore the human assessment analysis of these distortions. We tried to fill this lacuna by creating an S3D video dataset with diverse perceptual combinations of fog and haze artefacts. The dataset consists of 12 pristine and 360 distorted videos, and we used this dataset in our subjective study conducted by 24 observers. 
\subsection{Objective Quality Assessment}
\label{sec:objective}
Though 2D multimedia is still prevalent within the research community, but the advent of S3D content has made it more popular in the consumer digital world due to the immersive perceptual experience. To provide better QoE to the viewers, quality assessment of such S3D content has been a hot topic of research since many years now. Several authors~\cite{yasakethu2008quality, gorley2008stereoscopic,hewage2009quality,benoit2009quality,ma2016reorganized} have developed objective QA methods for S3D videos by performing 2D IQA and VQA models on left and right views of an S3D video. They concluded that VQA models offer better performance than IQA models, and involving depth features along with weighted pooling further enhance the performance. 

The cyclopean paradigm is one of the most efficient fusion techniques to incorporate depth information along with individual left and right scenes. Maalof~\textit{et al.}~\cite{maalouf2011cyclop} performed one of the first systematic frameworks to generate a cyclopean image from the 2D left and right views of an S3D image. They have compared the sensitivity coefficients of pristine and test cyclopean images to estimate the quality. Chen~\textit{et al.}~\cite{chen2013full} proposed a cyclopean image generation method based on binocular rivalry properties. They have applied 2D FR IQA model between pristine and distorted cyclopean images to estimate the quality of an S3D image. A similar cyclopean paradigm concept wss used by Battisti~\textit{et al.}~\cite{battisti2015perceptual} where they constructed the cyclopean frames for the pristine and test videos, and measured the binocular rivalry and depth features to estimate the quality of an S3D video. Boev~\textit{et al.}~\cite{1633754} proposed an FR S3D VQA metric based on computing cyclopean images, disparity maps and stereo similarity maps. Appina~\cite{appina2020complete} proposed an S3D IQA model based on performing the statistical analysis on the cyclopean image. We are inspired by the aforementioned cyclopean paradigm models~\cite{chen2013full,battisti2015perceptual,1633754,appina2020complete} and reused our previous work~\cite{appina2020complete} to generate cyclopean frames from the left and right views of an S3D video. 

Jiang~\textit{et al.}~\cite{jiang2018no} proposed an S3D NR VQA model based on performing the tensor decomposition of motion vector maps. They have calculated the statistical features and spectral entropies from tensor decomposition to estimate the quality of an S3D video. Several authors~\cite{Ha2011,solh2011no,zhang2016learning,han2015extended,sazzad2010spatio} proposed S3D VQA models based on computing the successive frame differences, temporal complexities, motion variations, statistical properties, etc. Hasan~\textit{et al.}~\cite{hasan2014no} measured an S3D video quality based on computing the edge strengths, energy errors and similarity scores of salient regions. Appina~\textit{et al.}~\cite{appina2018,appina2018no,appina2019subjective} proposed S3D FR and NR VQA models based on performing the statistical analysis on joint dependencies between temporal and depth features. Silva~\textit{et al.} and Mahamood~\textit{et al.}~\cite{silva2015no,mahmood2015objective} proposed S3D NR VQA methods based on measuring the structural properties of depth maps and correlation scores between the histograms of motion vector maps of an S3D video.

A supervised NR QA model for S3D videos is proposed by Yang~\textit{et al.}~\cite{91}. They have computed the saliency features and corresponding summation maps, and decomposed these features using sparse representation. Further, they have performed optical flow models to estimate the temporal features. Finally, Support Vector Regression (SVR) was performed on the combined spatial and temporal features to estimate the quality of an S3D video. Chen~\textit{et al.}~\cite{chen2018blind} proposed a supervised S3D NR VQA algorithm based on performing the auto-regression model on the features of disparity maps, left and right luminance summation and difference maps. Finally, they computed statistical measurements to estimate the quality of an S3D video. Ma~\textit{et al.}~\cite{ma2019stereoscopic} performed deep learning model on binocular fusion parameters of a stereo scene to estimate the quality of an S3D video. An end-to-end NR VQA model is developed by Feng~\textit{et al.}~\cite{100} which uses video blocks of the left and right views as inputs to extract spatiotemporal features via 3D convolution. The model consists of left and right channels along with a multi-stage growing attention channel, as well as fully connected (FC) and 3D convolution layers to predict the quality of S3D videos. 

The aforementioned objective QA models~\cite{yasakethu2008quality, gorley2008stereoscopic, hewage2009quality, benoit2009quality, ma2016reorganized, maalouf2011cyclop, chen2013full, battisti2015perceptual, 1633754, appina2020complete, jiang2018no, Ha2011, solh2011no, zhang2016learning, han2015extended, sazzad2010spatio, hasan2014no, appina2018, appina2018no, appina2019subjective, silva2015no, mahmood2015objective, 91, chen2018blind, ma2019stereoscopic, 100} utilize the performances of 2D IQA and VQA models, intra and inter scene characteristics, train and test sessions on features, etc. to estimate the quality of an S3D video. However, none of these models explore the NSS analysis on the binocular fused frames of an S3D video. We propose a completely blind and unsupervised (OU and DU) NR VQA model for S3D videos based on performing the NSS analysis at the patch level of cyclopean frames. We perform this analysis at multiple scales and orientations of spherical steerable pyramid decomposition and show the robust performance across datasets.
\begin{figure*}[!htbp]
\captionsetup[subfigure]{justification=centering}
\centering
\begin{subfigure}[b]{0.18\textwidth}
\includegraphics[width=3cm]{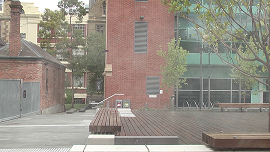}
\subcaption{\small $\#1$ (Fog ambiance), FADE = 1.3082.}
\label{fig:4ld0}
\end{subfigure}
\begin{subfigure}[b]{0.18\textwidth}
\includegraphics[width=3cm]{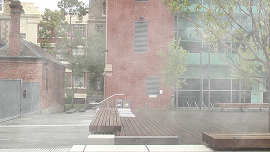}
\subcaption{\small $\#2$,\\ FADE = 2.1968.}
\label{fig:4ld1}
\end{subfigure}
\begin{subfigure}[b]{0.18\textwidth}
\includegraphics[width=3cm]{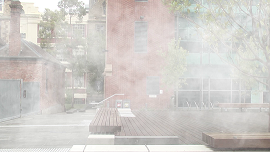}
\subcaption{\small $\#3$,\\ FADE = 3.4880.}
\label{fig:4ld2}
\end{subfigure}
\begin{subfigure}[b]{0.18\textwidth}
\includegraphics[width=3cm]{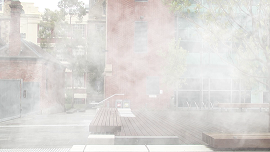}
\subcaption{\small $\#4$,\\ FADE = 4.3082.}
\label{fig:4ld3}
\end{subfigure}
\begin{subfigure}[b]{0.18\textwidth}
\includegraphics[width=3cm]{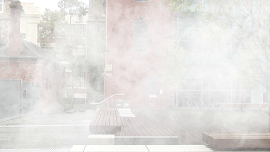}
\subcaption{\small $\#5$,\\ FADE = 5.2012.}
\label{fig:4ld4}
\end{subfigure}
\\
\begin{subfigure}[b]{0.18\textwidth}
\includegraphics[width=3cm]{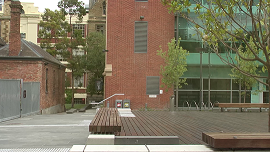}
\subcaption{\small $\#1$ (Haze ambiance),\\ FADE = 0.9324.}
\label{fig:4ld0_haze}
\end{subfigure}
\begin{subfigure}[b]{0.18\textwidth}
\includegraphics[width=3cm]{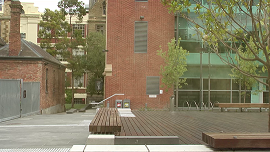}
\subcaption{\small $\#2$,\\ FADE = 1.9168.}
\label{fig:4ld1_haze}
\end{subfigure}
\begin{subfigure}[b]{0.18\textwidth}
\includegraphics[width=3cm]{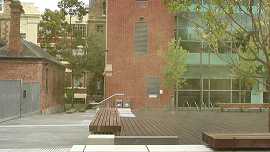}
\subcaption{\small $\#3$,\\ FADE = 2.9680.}
\label{fig:4ld2_haze}
\end{subfigure}
\begin{subfigure}[b]{0.18\textwidth}
\includegraphics[width=3cm]{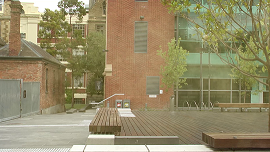}
\subcaption{\small $\#4$,\\ FADE = 3.6761.}
\label{fig:4ld3_haze}
\end{subfigure}
\begin{subfigure}[b]{0.18\textwidth}
\includegraphics[width=3cm]{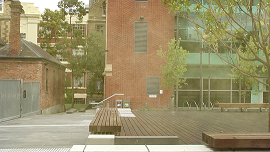}
\subcaption{\small $\#5$,\\ FADE = 4.3847.}
\label{fig:4ld4_haze}
\end{subfigure}
\caption{Illustration of $100^{th}$ frame from the left view of fog and haze distorted videos of Level 1 to Level 5. Fog Aware Density Evaluator (FADE) represents the visibility score of a scene.}
\label{fig:FogHazeLevels}
\end{figure*}
\section{Subjective Quality Assessment}
\label{sec:subjectivequalityasessment}
This section describes the process of designing and executing the proposed stereoscopic video dataset. It also discusses the experimental setup that was used for performing the subjective study experiment on the proposed dataset.
\subsection{Reference Video Sequences}
We selected our pristine S3D video sequences from three well-known stereoscopic video datasets:  DML-ITRACK-3D video dataset~\cite{fang2017visual}, MMSPG-EPFL S3D video dataset~\cite{10.1117/12.839438} and Mobile 3D TV - stereo video dataset~\cite{mobile3dtv}. 
\subsubsection{DML-ITRACK-3D video dataset}
This dataset consists of 47 pristine S3D videos. The video sequences have true HD resolution with a frame rate of 25 fps and a duration of either 10 sec. or 16 sec. All video sequences are encoded in YUV 422P 10-bit format and saved in .mov container. 
\subsubsection{MMSPG-EPFL S3D video dataset}
This dataset is composed of 6 uncompressed S3D video sequences with a combination of four indoor and two outdoor scenes. The video sequences have a resolution of $1920 \times 1080$ and a duration of 10 sec. with a frame rate of 25 fps. The video sequences are encoded in YUV 420P format and saved in .avi container.
\subsubsection{Mobile 3D TV - stereo video dataset}
This dataset has 28 reference S3D video sequences with true HD resolution and multiple durations and frame rates. All video sequences are encoded in YUV 420P format and saved in .avi container.

We are motivated by~\cite{pourazad2011effect,pinson2013selecting} to perform a preliminary subjective study on the uncompressed videos of the aforementioned datasets to choose a pristine S3D videos representative set. We have involved 7 subjects in this study who were asked to rate each video on a scale of 1 to 5 based on their perceptual feel of the video quality in terms of spatial information, temporal activity, depth, and general feeling. A rating of 1 indicates `Bad', 2 indicates `Poor', 3 indicates `Fair', 4 indicates `Good' and 5 indicates `Excellent'. Based on the received feedback and subjective score agreement, we have selected 9 videos from DML-ITRACK-3D video dataset, 2 videos from MMSPG-EPFL S3D video dataset and 1 video from Mobile 3D TV - stereo video dataset, and each video is trimmed to 10 sec. duration. Further, we have converted pristine video codec formats to the YUV 420P 8 bit format in order to ensure smooth playback on the 3D TV.  Fig.~\ref{fig:all reference video frames} shows the $100^{th}$ frame from the left view of 12 pristine videos of our dataset.

Figs.~\ref{fig:SITI} and \ref{fig:DSIDTI} show the variation of Spatial Index (SI) and Temporal Index (TI) scores, and disparity Spatial Index (DSI) and disparity Temporal Index (DTI) scores of pristine video sequences of the dataset~\cite{de2013toward}. We have computed the mean score of SI and TI scores of left and right views to represent the overall SI and TI scores of an S3D video. We derived DSI and DTI scores from disparity maps of pristine S3D videos using SSIM based stereo matching algorithm. It is clearly evident from the plots that the selected reference videos contain a wide range of spatial, temporal, and disparity information. The 12 videos are plotted by assigning alphabet labels ranging from a-l corresponding to the labels used in Fig.~\ref{fig:all reference video frames}.
\subsection{Test Video Sequences}
We have created 360 distorted video sequences by adding the fog and haze ambiance levels on the left and right views of an S3D video. The ambiance levels were chosen to cover the wide range of perceptual qualities. 
\subsubsection{Fog distortion}
Fog distortion is a kind of opaque obscurity which is formed due to water droplets suspended in the air. We have used Adobe Premiere Pro 2020~\cite{adobe2020} software to simulate fog distortion on the videos. We have applied 5 perceptual ambiance levels on left and right views of the 12 reference S3D videos to generate 180 fog distorted videos with a combination of 60 symmetrically and 120 asymmetrically distorted video sequences.
\subsubsection{Haze distortion}
The haze distortion is caused by dry particles, dust, smoke, etc. suspended in the air. Similar to the fog distortion, each pristine left and right video was distorted at 5 perceptual haze ambiance levels by utilizing the Adobe Premiere Pro 2020 software. A total of 180 distorted S3D videos have been created comprising of 60 symmetrically and 120 asymmetrically distorted video sequences.

Fig.~\ref{fig:FogHazeLevels} shows the $100^{th}$ frame from the left view of fog and haze distorted videos at different perceptual ambiances with the corresponding Fog Aware Density Evaluator (FADE)~\cite{7159086} scores. It is clear from the plot that the perceived quality of a video varies with the ambiance level and also indicates that the proposed dataset contains a wide range of ambiance levels.
\subsection{Subjective Study Experiment}
\label{sec:substudy exp}
The subjective study was performed in the Lab for Video and Image Analysis (LFOVIA) at the Indian Institute of Technology Hyderabad (IITH). We used the LG Television (LG49UF850T) to display the videos. The display is a circularly polarized TV with a screen resolution of ultra HD $(3840 \times 2160)$ and it performs 3D projection based on Film-type patterned retarder (FPR) technology. We strictly adhered to the recommendations of ITU-R~\cite{union2015subjective} to setup the other subjective study settings. 
\begin{figure}[!htbp]
\centering
\begin{subfigure}[b]{0.24\textwidth}
\includegraphics[height=3cm,width=4cm]{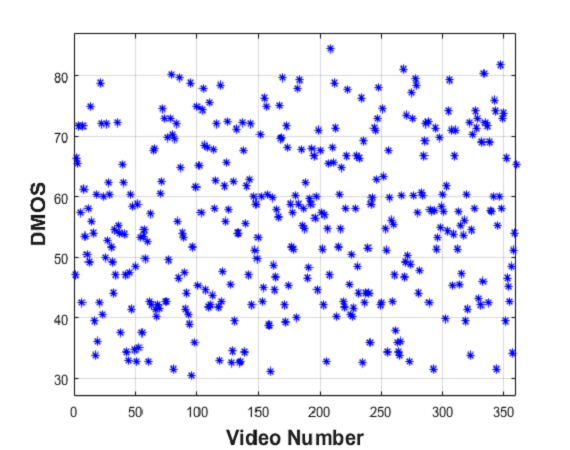}
\subcaption{\small DMOS scores distribution.}
\label{fig:DMOS_scatter}
\end{subfigure}
\begin{subfigure}[b]{0.24\textwidth}
\includegraphics[height=3cm,width=4cm]{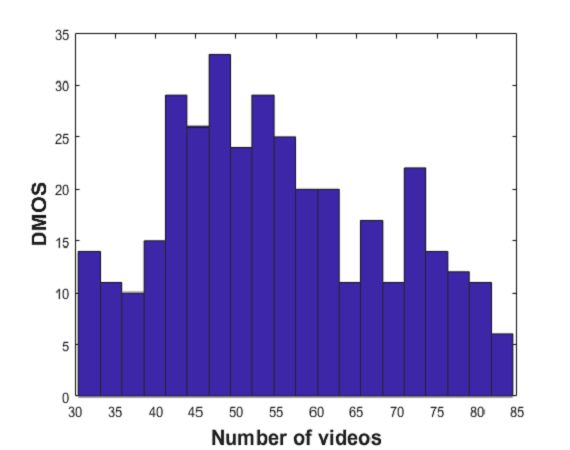}
\subcaption{\small Histogram of DMOS scores.}
\label{fig:DMOS_hist}
\end{subfigure}
\caption{Illustration of DMOS scores variation.}
\label{fig:DMOS}
\end{figure}

We have included 24 naive and amateur participants to perform the subjective study. The gender distribution of observers is not limited and the average age of all participants is 24 years. We have conducted a 3-minute demo session for all participants to familiarize them with 3D videos and test stimuli perceptual variations. We also requested their feedback on the visual discomfort experience, and none of the participants reported any nausea or fatigue during the demo session. 
\begin{table}[!htbp]
\small
\caption{\small Mean ($\mu$), median ($m$) and standard deviation ($\sigma$) of LCC and SROCC over 100 inter subject trials across different distortions of the proposed dataset.}
\centering 
\begin{tabular}{|c| c| c| c| c| c| c|} 
\hline
  \multirow{2}{*}{Score} & \multicolumn{2}{c|}{Fog}& \multicolumn{2}{c|}{Haze}& \multicolumn{2}{c|}{Overall dataset}\\
 \cline{2-7} 
  &  {LCC}  & {SROCC} & {LCC} &{SROCC}  & {LCC} & {SROCC}  \\
\hline
$\mu$& 0.955&0.952&0.942&0.944&0.947&0.945 \\
\hline
$m$& 0.957&0.954&0.943&0.945&0.959&0.943 \\
\hline
$\sigma$& 0.048&0.048&0.050&0.049&0.049&0.049  \\
\hline
\end{tabular}
\label{table:internal}
\end{table}

The subjective experiment is conducted in three sessions and each session duration is 30 minutes approximately. We have provided at least 24 hours of break between each session of each participant to allow them to recover from any type of visual fatigue suffered during the study. We have relied on hardware renderers to playback videos smoothly on display, and there is no ability to interact with the graphic user interface. Due to this, participants were instructed to call out the perceptual quality score after watching the video. The video sequences were arranged in a random order of perceptual ambiance levels of fog and haze distortions without repetition. We have performed the subjective assessment according to Single Stimulus method and Absolute Category Scale (ACR) with the hidden reference (HR) protocol. The subjects did the rating on video quality based on five levels such as `Bad,' `Poor,' `Fair,' `Good,' or `Excellent'.
\subsection{Subjective Score Analysis}
\label{subjective score analysis}
In the subjective study, we have collected 372 (12 reference + 180 fog + 180 haze) ratings from each subject. We followed the procedure recommended by ITU-R~\cite{union2015subjective} to process the subjective scores. We first calculated the difference score between the quality scores of a test video and the corresponding pristine video of the same subject. 
\begin{align}
    \Delta_{ij} = V_{{ref}_{ij}} - V_{{dist}_{ij}},
\label{Eq1}
\end{align}
where $i$ and $j$ indicate the subject and video sequence id. $V_{{ref}_{ij}}$ and $V_{{dist}_{ij}}$ represent the assessment scores of pristine and test videos by the same subject. $\Delta_{ij}$ is the difference score between $V_{{ref}_{ij}}$ and $V_{{dist}_{ij}}$. 

We have performed the normalization by computing the mean $(\mu_{i})$ and standard deviation $(\sigma_{i})$ scores from difference scores.
\begin{align}
N_{ij} = \frac{\Delta_{ij} - \mu_{i}}{\sigma_{i}},
\label{Eq2}
\end{align}
where $N_{ij}$ represents the normalized scores. We followed the procedure recommended by \cite{union2015subjective} for subject rejection, and we found no outliers. The $N_{ij}$ scores were present in the range of $[-3, 3]$ and we scaled them to $[0, 100]$ to calculate $N^{'}_{ij}$ scores.
\begin{align}
N^{'}_{ij} = \frac{(N_{ij}+3) \times 100}{6},
\label{Eq3}
\end{align}
\begin{figure*}[!htbp]%
\captionsetup[subfigure]{justification=centering}
\centering
\begin{subfigure}[b]{0.27\textwidth}
\includegraphics[width=4.5cm,height=3cm]{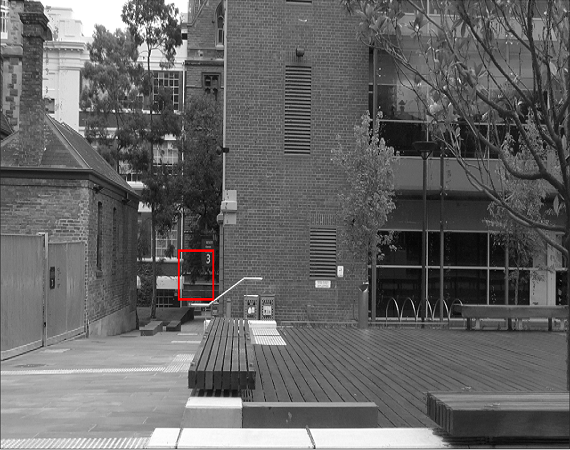}
\subcaption{\small Pristine cyclopean frame.}
\label{fig:PristCyclop} 
\end{subfigure}
\begin{subfigure}[b]{0.27\textwidth}
\includegraphics[width=4.5cm,height=3cm]{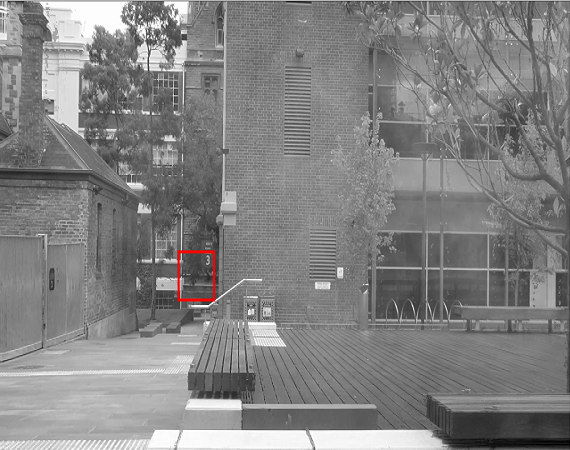}
\subcaption{\small Fog cyclopean frame.} 
\label{fig:FogCyclop}
\end{subfigure}
\begin{subfigure}[b]{0.27\textwidth}
\includegraphics[width=4.5cm,height=3cm]{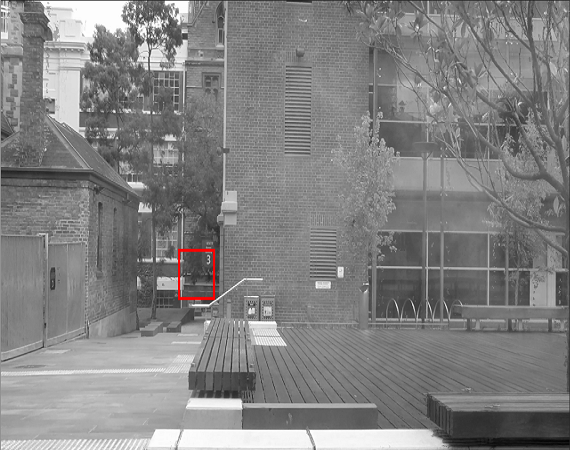}
\subcaption{\small Haze cyclopean frame.} 
\label{fig:HazeCyclop}
\end{subfigure}
\\
\begin{subfigure}[b]{0.27\textwidth}
\includegraphics[width=4.5cm,height=3cm]{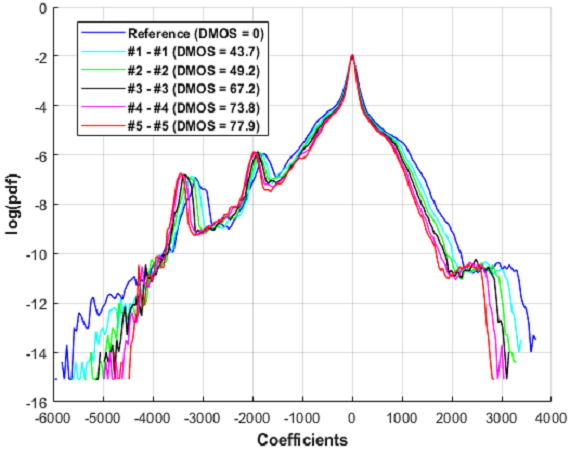}
\subcaption{\small Log - histograms of pristine and symmetric fog distorted versions computed at $\theta = 0 ^{\circ}$ and $\Phi = -90 ^{\circ}$.}
\label{fig:SymFogPhiN90} 
\end{subfigure}
\begin{subfigure}[b]{0.27\textwidth}
\includegraphics[width=4.5cm,height=3cm]{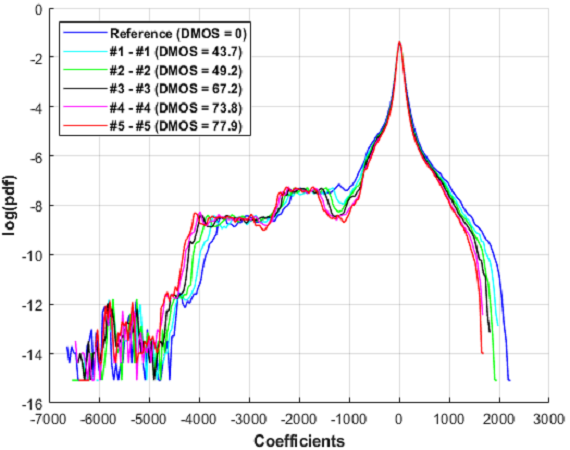}
\subcaption{\small Log - histograms of pristine and symmetric fog distorted versions computed at $\theta = 0 ^{\circ}$ and $\Phi = 0 ^{\circ}$.} 
\label{fig:SymFogPhi0}
\end{subfigure}
\begin{subfigure}[b]{0.27\textwidth}
\includegraphics[width=4.5cm,height=3cm]{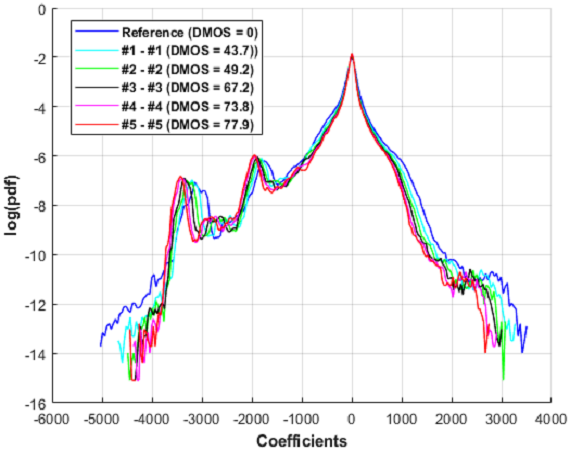}
\subcaption{\small Log - histograms of pristine and symmetric fog distorted versions computed at $\theta = 0 ^{\circ}$ and $\Phi = 90 ^{\circ}$.} 
\label{fig:SymFogPhi90}
\end{subfigure}
\\
\begin{subfigure}[b]{0.27\textwidth}
\includegraphics[width=4.5cm,height=3cm]{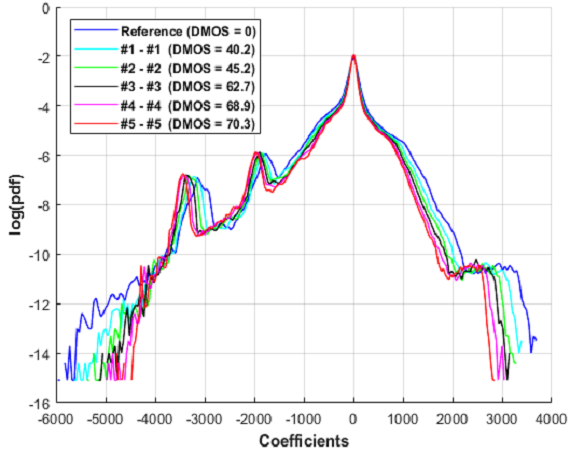}
\subcaption{\small Log - histograms of pristine and symmetric haze distorted versions computed at $\theta = 0 ^{\circ}$ and $\Phi = -90 ^{\circ}$.}
\label{fig:SymHazePhiN90} 
\end{subfigure}
\begin{subfigure}[b]{0.27\textwidth}
\includegraphics[width=4.5cm,height=3cm]{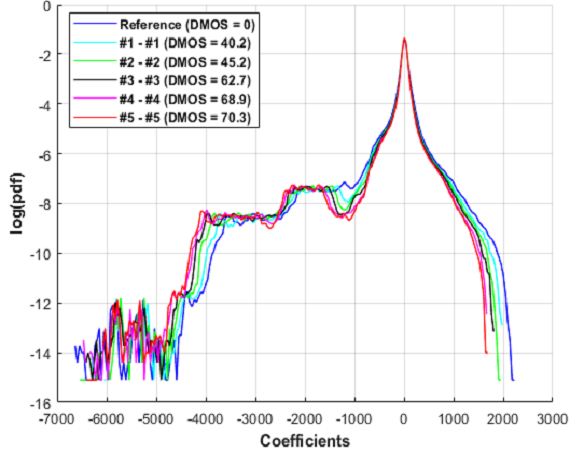}
\subcaption{\small Log - histograms of pristine and symmetric haze distorted versions computed at $\theta = 0 ^{\circ}$ and $\Phi = 0 ^{\circ}$.} 
\label{fig:SymHazePhi0}
\end{subfigure}
\begin{subfigure}[b]{0.27\textwidth}
\includegraphics[width=4.5cm,height=3cm]{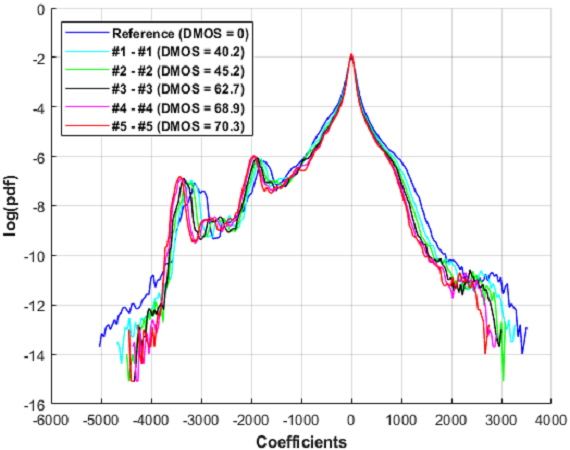}
\subcaption{\small Log - histograms of pristine and symmetric haze distorted versions computed at $\theta = 0 ^{\circ}$ and $\Phi = 90 ^{\circ}$.} 
\label{fig:SymHazePhi90}
\end{subfigure}
\caption{Illustration of log-histograms of pristine and corresponding symmetrically distorted versions of fog and haze ambiance videos computed at first scale, $\theta = 0 ^{\circ}$ and $\Phi = (-90 ^{\circ}, 0 ^{\circ}, 90 ^{\circ}$) orientations.}
\label{fig:CYclopFeatSymm}
\end{figure*}

Finally, we computed the Difference Mean Opinion Scores (DMOS) by calculating the mean score across $N^{'}_{ij}$ ratings of a video.
\begin{align}
\text{DMOS}_{j}=\frac{\sum\limits_{{i}=1}^{Z} N^{'}_{ij}} {Z},
\label{Eq11}
\end{align}
where $Z ( = 24) $ represents the total number of subjects.

Figs.~\ref{fig:DMOS_scatter} and \ref{fig:DMOS_hist} show the DMOS distribution and histogram of test videos of the dataset, respectively. It is evident from the plots that the dataset has a wide range of perceptual qualities.

Table~\ref{table:internal} shows the efficacy of the subjective experiment by analyzing the internal structure of the dataset. We have randomly divided the DMOS scores into two halves without overlap, and computed the Linear Correlation Coefficient (LCC) and Spearman's Rank Order Correlation Coefficient (SROCC) between the randomly divided two sets. We repeated the analysis 100 times and computed mean ($\mu$), median ($m$) and standard deviation ($\sigma$) of LCC and SROCC scores of 100 trials. It is clearly evident from these numbers that the subjects agreed on the video ratings.
\section{Objective Quality Assessment}
\label{sec:objectivequalityasessment}
Several psychovisual experiments~\cite{ingle1982analysis,goodale1992separate,deangelis1999organization} were conducted on the macaque's central nervous system to explore the neural functionalities and ventral-dorsal architecture. They concluded that the ventral and dorsal streams have simple and complex cells, and both streams are sensitive to high spatial and temporal frequencies and responsible for processing the depth information. Inspired by these psychovisual experiments, several research works~\cite{potetz2003statistical,liu2011statistical,khan2015full} explored the marginal statistics of stereoscopic scene components. They concluded that the shape of the histograms of subband coefficients of stereo scene components have heavy tails and sharp peaks, and these histograms can be accurately modeled with Univariate Generalized Gaussian Distribution (UGGD). The combined results of psychovisual experiments and the successful application of statistical models motivated us to study the stereoscopic 3D video level scene statistics by performing the spherical steerable pyramid decomposition at multiple scales and multiple orientations. 
\begin{figure*}[!htbp]%
\captionsetup[subfigure]{justification=centering}
\centering
\begin{subfigure}[b]{0.245\textwidth}
\includegraphics[width=4.5cm,height=2.5cm]{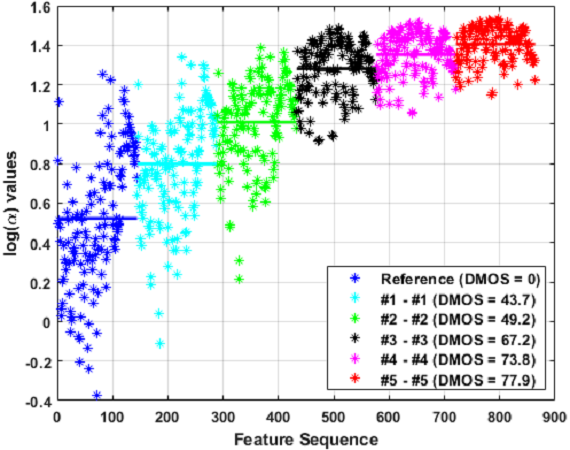}
\subcaption{\small $\alpha$ feature distribution of symmetrically distorted fog S3D videos.}
\label{fig:FogSymmPFeat} 
\end{subfigure}
\begin{subfigure}[b]{0.245\textwidth}
\includegraphics[width=4.5cm,height=2.5cm]{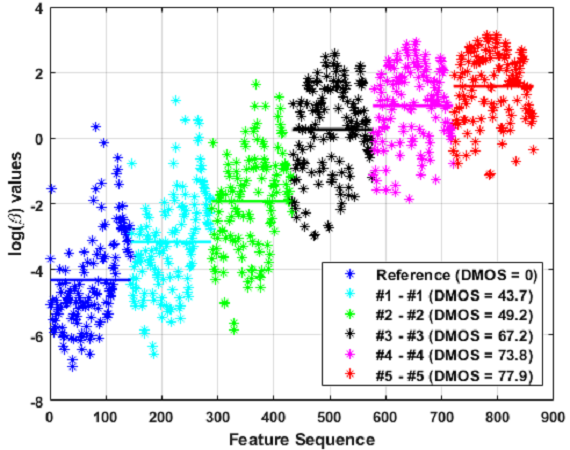}
\subcaption{\small $\beta$ feature distribution of symmetrically distorted fog S3D videos.} 
\label{fig:FogSymmSFeat}
\end{subfigure}
\begin{subfigure}[b]{0.245\textwidth}
\includegraphics[width=4.5cm,height=2.5cm]{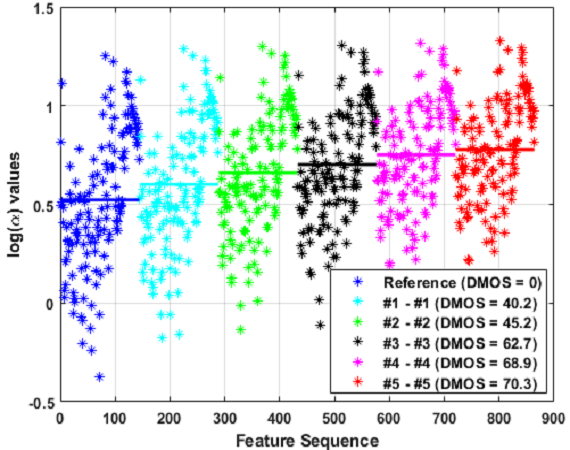}
\subcaption{\small$\alpha$ feature distribution of symmetrically distorted haze S3D videos.} 
\label{fig:HazeSymmPFeat}
\end{subfigure}
\begin{subfigure}[b]{0.245\textwidth}
\includegraphics[width=4.5cm,height=2.5cm]{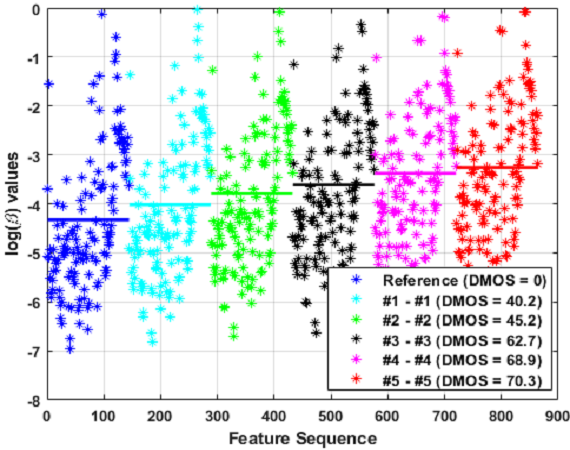}
\subcaption{\small $\beta$ feature distribution of symmetrically distorted haze S3D videos.} 
\label{fig:HazeSymmSFeat}
\end{subfigure}
\caption{Visualization of UGGD $(\alpha, \beta)$ feature distribution of pristine and corresponding symmetrically distorted versions of fog and haze S3D videos.}
\label{fig:PSFeatSetScatter}
\end{figure*}
\subsection{Proposed Algorithm}
The proposed algorithm consists of four stages. In the first stage, we generate a video from cyclopean frames of left and right views of an S3D video. The second performs the spherical steerable pyramid decomposition on the produced cyclopean video at multiple scales and multiple orientations. In the third stage, we explain the NSS analysis on S3D videos. The fourth and final stage performs the CBSE score computation of an S3D video. 
\subsubsection{Cyclopean Frame}
Our HVS is designed to perform a virtual fusion of the left and right retinal views perceived by the human eyes into a single view using locally matched regions of the stereoscopic pair. Levelt proposes a linear combination model to explain binocular rivalry in cyclopean images~\cite{levelt1968binocular}. The study stated that binocular rivalry occurs due to differences in the strength of the retinal stimulus of the stereo pair. Another study by Messai~\textit{et al.}~\cite{MESSAI2020115772} found that using cyclopean images provided more accurate results in the case of quality assessment of S3D scenes. The aforementioned results motivated us to generate a cyclopean image from the left and right scenes of an S3D video. 

We followed our previous work~\cite{appina2020complete} to construct cyclopean frames of an S3D video. The formula used is as follows:
\begin{equation} 
C(x,y) = W_{L}(x,y) \times I_{L}(x,y) + W_{R}(x+d,y) \times I_{R}(x+d,y),
\label{eq:coher}
\end{equation}
where $C$ indicates the cyclopean frame. $I_{L}$ and $I_{R}$ represent left and right frames, respectively. Spatial coordinates are ($x, y$). $d$ represents the disparity pixel computed from the stereo correspondence between $I_{L}$ and $I_{R}$ based on SSIM based stereo matching algorithm~\cite{chen2013full}. $W_{L}$ and $W_{R}$ are weights assigned to left and right views of an S3D scene. The weights $W_{L}$ and $W_{R}$ are computed as follows:
\begin{align}
    W_{L}(x,y) &= \frac{rms(S_L(x,y))}{rms(S_L(x,y))+rms(S_R(x+d,y))},\\
    W_{R}(x,y) &= \frac{rms(S_R(x+d,y))}{rms(S_L(x,y))+rms(S_R(x+d,y))},
\end{align}
where $rms$ indicates the root mean square value. $S_L$ and $S_R$ represent saliency maps of $I_{L}(x,y)$ and $I_{R}(x+d,y)$ computed based on Graph-Based Visual Saliency (GBVS) model~\cite{harel2007graph}. These weights help us to understand the symmetric and asymmetric nature of a stereoscopic scene. We partitioned the generated cyclopean frames into $X \times Y \times T $ nonoverlapping blocks and empirically selected $X = 120$ and $Y = 120$ based on the best performance analysis. $T$ is the total number of frames in a video.
\subsubsection{Spherical Steerable Pyramid Decomposition}
\label{spherical steerable filter}
The spherical steerable filter is used for adaptive steering of three-dimensional axially symmetric functions~\cite{93808, 1530451}. If $f(x,y,t)$ represents the function with the axis of rotational symmetry pointing towards the $a, b, c$ direction cosines, and if we perform 3D rotation of this function by a transformation $\Theta$, then the rotated function is written as follows,
\begin{equation}
    f^{\Theta}(x,y,t)= \delta(r)Q_{M}(m),
\end{equation}
where $r=\sqrt{x^{2}+y^{2}+t^{2}}$ and $\delta(r)$ is a three dimensional spherically symmetric windowing function. $m = ax + by + ct$ and $Q_{M}(m)$ is an $M^{th}$ order degree polynomial in $m$. The filter orientations are represented as the direction cosines of the axis of symmetry and expressed in spherical coordinates as follows:
\begin{align}
a = cos(\theta) sin(\Phi), b &= sin(\theta) sin(\Phi), c = cos(\Phi), 
\end{align}
We are inspired by \cite{Somdyuti2017} to perform the aforementioned subband decomposition at three spatial scales and multiple azimuth angles $\theta$ $( = 0^{\circ}, 45^{\circ}, 90^{\circ}, 135^{\circ}, 180^{\circ}, 225^{\circ}, 270^{\circ}, 315^{\circ}, 360^{\circ})$ and elevation angles $\Phi$ $( = -90^{\circ}, -45^{\circ},  0^{\circ}, 45^{\circ}, 90^{\circ})$.
\subsubsection{NSS based feature computation}
In recent years, several researchers \cite{khan2015full,su2013color,lee2016toward} have been inspired by the work of Liu \textit{et al.} \cite{liu2011statistical} to study the NSS characteristics of 2D and 3D images and videos. 
\begin{align}
\gamma({\bf{z}}|\alpha,\beta)&=\left[\frac{\alpha \times \Omega(\alpha,\beta)}{2 \times \Gamma(1/\alpha)}\right] exp[-(z \times \Omega(\alpha,\beta))^{\alpha}],\\
\Omega(\alpha,\beta)&={\beta^{-1}}\left[\frac{\Gamma(3/\alpha)}{\Gamma(1/\alpha)}\right]^{1/2},
\label{eq:uggd}
\end{align}
where $\alpha$ and $\beta$ are UGGD model fitting coefficients and these parameters indicate the shape and spread of the distribution. $\bf{z}$ is a random vector and $\Gamma(.)$ is a gamma function. We compute $\alpha$ and $\beta$ parameters at the aforementioned spatial scales and orientations of spherical steerable pyramid decomposition. 
\begin{figure}[!htbp]%
\captionsetup[subfigure]{justification=centering}
\centering
\begin{subfigure}[b]{0.225\textwidth}
\includegraphics[height=2.7cm]{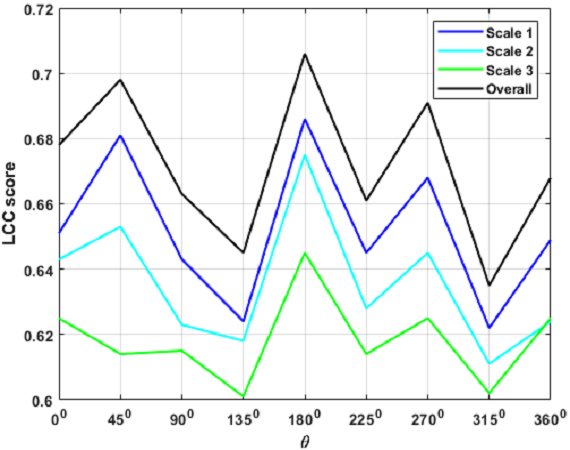}
\subcaption{\small $\theta$.}
\label{fig:PerTheta} 
\end{subfigure}
\begin{subfigure}[b]{0.225\textwidth}
\includegraphics[height=2.7cm]{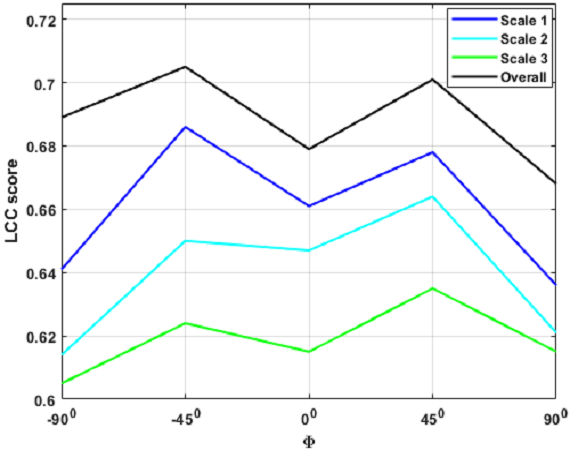}
\subcaption{\small $\Phi$.} 
\label{fig:PerPhi}
\end{subfigure}
\caption{Illustration of LCC score variation of CBSE model computed at different $\theta$ and $\Phi$ orientations and scales.}
\label{fig:OriSetScatter}
\end{figure}
\vspace{-0cm}

Fig.~\ref{fig:PristCyclop} shows the $100^{th}$ cyclopean frame generated from the left and right views of pristine `RMIT University Courtyard' S3D video sequence, the red color border indicates a nonoverlapping patch. Figs.~\ref{fig:FogCyclop} and \ref{fig:HazeCyclop} show the cyclopean frames of fog and haze distorted versions of the same pristine video, respectively. The figures clearly demonstrate perceptual quality variations with respect to the type and level of distortion. Figs.~\ref{fig:SymFogPhiN90}, \ref{fig:SymFogPhi0}, \ref{fig:SymFogPhi90}, \ref{fig:SymHazePhiN90}, \ref{fig:SymHazePhi0} and \ref{fig:SymHazePhi90} show the log histograms of nonoverlapping patch of the pristine and corresponding symmetrically distorted versions of fog and haze ambiances, respectively. From these plots, it is possible to make two important observations: 1) sharp peaks along with heavy tails are evident in the histograms, and therefore GGD models can be endorsed, and, 2) there is significant variation among the histograms with respect to the type and level of distortion. The aforementioned histograms are computed at first scale, $\theta = 0 ^{\circ}$ and $\Phi = (-90 ^{\circ}, 0 ^{\circ}, 90 ^{\circ})$ orientations, and we observe the similar trend across all plots. These findings have motivated us to use UGGD to model the subband coefficients. 
\begin{table*}[htbp]
\small
\caption{Off-the-shelf 2D and 3D IQA/VQA models and proposed algorithm performance evaluation in terms of LCC, SROCC and RMSE on the symmetrically and asymmetrically distorted versions of the VAD stereo video dataset.} 
\centering 
\begin{tabular}{|c|c| c| c| c| c| c| c| c| c| c|} 
\hline
\multirow{2}{*}{ {Model Type}}&  \multirow{2}{*}{{{Algorithm}}} & \multicolumn{3}{c|}{{{Symmetric}}}& \multicolumn{3}{c|}{{{Asymmetric}}}& \multicolumn{3}{c|}{{{Overall}}}\\
 \cline{3-11} 
&  &  {{{LCC}}}  & {{{SROCC}}} & {{{RMSE}}} &{{{LCC}}}  & {{{SROCC}}} & {{{RMSE}}}  &{{{LCC}}}  & {{{SROCC}}} & {{{RMSE}}} \\
\hline
\multirow{3}{*}{{{2D FR IQA}}} &{{SSIM  \cite{wang2004ssim}}} & 0.425  & 0.399 & 14.696&0.408 & 0.388&14.706&0.417 &0.391&14.747  \\
\cline{2-11}
&{{MS-SSIM  \cite{wang2003multiscale}}} & 0.458 & 0.407 & 14.600  &0.433	&0.371	&14.576 &0.439	&0.402	&14.637\\
\cline{2-11}
&{{VIF \cite{Sheikh2006}}}& 0.568  &0.548 & 13.285  &0.480&0.430	&14.126 &0.498	&0.435	&13.523\\
\hline
\hline
\multirow{4}{*}{{{2D NR IQA}}}& {{BRISQUE \cite{mittal2012no}}} & 0.251 &0.184&15.628&0.185&0.219&16.093&0.178&0.199&16.209\\
\cline{2-11}
&{{NIQE \cite{mittal2013making}}}& 0.421 & 0.352 & 14.379  &0.327&0.294	&15.693 &0.380	&0.356	&15.649\\
\cline{2-11}
&{{PIQUE \cite{venkatanath2015blind}}}& 0.231 & 0.209 & 15.739  &0.216&0.208&15.765 &0.219	&0.206	&15.090\\
\cline{2-11}
&{{ILNIQE \cite{zhang2015feature}}}& 0.469 & 0.459 & 14.255  &0.323&0.241&15.406 &0.374	&0.319	&14.961\\
\hline
\hline
\multirow{2}{*}{{{2D FR VQA}}}&{{STMAD \cite{vu2011spatiotemporal}}} & 0.754 & 0.686 & 8.810 & 0.621&0.589	&11.171 & 0.677&0.628&10.424\\
\cline{2-11}
&{{SpEED \cite{7979533}}}  & 0.757 & 0.700 & 8.809 & 0.715& 0.696 &9.310&0.749&0.700&9.558\\%
\hline
\hline
\multirow{2}{*}{{{2D NR VQA}}}&{{VIIDEO \cite{mittal2015completely}}} & 0.262 & 0.234 & 15.582 & 0.286&0.271	&15.758 & 0.272&0.255&15.895\\
\cline{2-11}
&{{NSTSS \cite{9059006}}}  & 0.418 & 0.289 & 14.766 & 0.396& 0.278 &14.829&0.395&0.283&14.830\\%
\hline
\hline
\multirow{2}{*}{{{3D FR IQA}}}&{{STRIQE \cite{khan2015full}}} & 0.601 & 0.602 & 12.894 & 0.566	&0.534	&11.814 & 0.577&0.569&11.773\\
\cline{2-11}
&{{MJ3DQA \cite{chen2013full}}}  &  0.654 & 0.624 & 10.955 & 0.589	&0.491	&11.958
& 0.617&0.581&10.823\\
\hline
\hline
\multirow{1}{*}{{{3D NR IQA}}}&{{MO-NIQE \cite{appina2020complete}}}  & 0.229 &0.200  &15.973  &0.165 &0.136  &16.319 &0.213 &0.193 &15.085 \\%
\hline
\hline
\multirow{1}{*}{{{3D FR VQA}}}&{{$\text{DeMo}_{3D}$ \cite{appina2018}}}  & 0.753 & 0.665 & 8.811 & 0.625	&0.610	&11.169 & 0.680&0.641&10.422\\
\hline
\hline
\multirow{3}{*}{{{3D NR VQA}}}&{{VQUEMODES \cite{appina2018no}}}   & 0.873 & 0.834 & 3.166 & 0.846& 0.833 &5.398&0.845&0.849&5.239\\%
\cline{2-11}
&{{$\text{MoDi}_{3D}$ \cite{appina2019subjective}}} & 0.486 & 0.447 & 14.148  & 0.336&	0.302	&15.411 & 0.407&0.632	&14.716\\
\cline{2-11}
&{{CBSE}} & 0.767 & 0.749 & 8.750  & 0.663&0.643&10.610 & 0.717&0.685	&9.805\\
\hline
\end{tabular}
\label{table:VADStereo}
\end{table*}
\begin{table*}[htbp]
\caption{Performance evaluation (LCC, SROCC and RMSE) of off-the-shelf 2D \& 3D IQA and VQA models on the symmetrically and asymmetrically distorted versions of the fog and haze S3D videos.} 
\centering 
\begin{tabular}{|c| c| c| c| c| c| c| c| c| c| c| c| c| } 
\hline
\multirow{3}{*}{Model} &  \multicolumn{6}{c|}{ Fog videos} &  \multicolumn{6}{c|}{ Haze videos}\\
\cline{2-13}
  &  \multicolumn{3}{c|}{ Symm}  & \multicolumn{3}{c|}{ Asymm}  &\multicolumn{3}{c|}{ Symm}  & \multicolumn{3}{c|}{ Asymm} \\
  \cline{2-13}
 &  LCC &  SROCC & RMSE &  LCC &  SROCC &  RMSE & LCC &  SROCC &  RMSE &  LCC &  SROCC &  RMSE  \\
 \hline
 SSIM\cite{wang2004ssim}&0.440&0.423&14.687&0.437&0.384&14.609&0.452&0.421&14.614&0.410&0.400&14.889\\
\hline
 MS-SSIM\cite{wang2003multiscale}&0.464&0.409&14.540&0.438&0.398&14.453&0.486&0.470&14.1884&0.456&0.448&14.020\\
\hline
{{VIF \cite{Sheikh2006}}}&0.538&0.521&11.974&0.300&0.281&15.578&0.558&0.522&11.707&0.396&0.337&15.102\\
\hline
{{BRISQUE \cite{mittal2012no}}}&0.313&0.337&15.491&0.265&0.215&15.895&0.332&0.333&15.402&0.340&0.315&15.167\\
\hline
{{NIQE \cite{mittal2013making}}}&0.544&0.463&11.919&0.370&0.345&14.988&0.480&0.411&14.202&0.369&0.325&15.140\\
\hline
{{PIQUE \cite{venkatanath2015blind}}}&0.268&0.207&15.896&0.196&0.117&15.627&0.178&0.154&	16.670&0.176&0.170&16.089\\
\hline
{{ILNIQE \cite{zhang2015feature}}}&0.486&0.445&14.216&0.366&0.325&14.909&0.477&0.455&14.373&0.293&0.254&15.729\\
\hline
{{STMAD \cite{vu2011spatiotemporal}}}&0.761&0.695&8.602&0.659&0.602&10.235&0.611&0.591&11.394&0.550&0.494&13.553\\
\hline
{{SpEED \cite{7979533}}}&0.760&0.696&8.600&0.697&0.627&9.975&0.636&0.544&	11.187&0.592&0.547&13.179\\
\hline
{{VIIDEO \cite{mittal2015completely}}}&0.234&0.125&15.824&0.156&0.176&16.185&0.150&	0.109&16.76&0.175&0.130&16.089\\
\hline
{{NSTSS \cite{9059006}}}&0.521&0.395&12.118&0.421&0.315&14.694&0.488&0.434&14.138&0.358&0.326&15.149\\
\hline
{{STRIQE \cite{khan2015full}}}&0.675&0.638&9.751&0.577&0.595&13.352&0.656&	0.623&11.023&0.583&0.544&13.374\\
\hline
{{MJ3DQA \cite{chen2013full}}}&0.693&0.683&10.072&0.628&0.575&10.796&0.731&	0.712&9.899&0.708&0.704&10.015\\
\hline
{{MO-NIQE \cite{appina2020complete}}}&0.242&0.225&16.131&0.185&0.152&15.910&0.191&0.174&16.309&0.120&0.115&16.350\\
\hline
{{$\text{DeMo}_{3D}$ \cite{appina2018}}}&0.759&0.710&8.602&0.700&0.686&9.973&0.645&0.582&11.180&0.602&0.571&13.166\\
\hline
{{VQUEMODES \cite{appina2018no}}} &0.876&0.854&3.033&0.866&0.857&4.993&0.850&0.828&4.662&0.791&0.7501&5.604\\
\hline
{{$\text{MoDi}_{3D}$ \cite{appina2019subjective}}} &0.573&0.560&11.633&0.370&0.359&14.824&0.454&0.420&14.581&0.381&0.345&15.612\\
\hline
{{CBSE}} &0.798&0.769&8.413&0.696&0.657&9.876&0.728&0.728&9.401&0.653&0.611&10.280\\
\hline
\end{tabular}
\label{table:VADStereoEach}
\end{table*}
\begin{table*}[htbp]
\small
\caption{Performance evaluation (LCC, SROCC and RMSE) of off-the-shelf 2D and 3D IQA and VQA models and proposed CBSE algorithm on IRCCYN, LFOVIAPh1 and LFOVIAPh2 S3D video datasets.} 
\centering 
\begin{tabular}{|c|c| c| c| c| c| c| c| c| c| c|} 
\hline
Model &  \multirow{2}{*}{Algorithm} & \multicolumn{3}{c|}{IRCCYN}& \multicolumn{3}{c|}{LFOVIAPh1}& \multicolumn{3}{c|}{LFOVIAPh2}\\
 \cline{3-11} 
&  &  {LCC}  & {SROCC} & {RMSE} &{LCC}  & {SROCC} & {RMSE}  &{LCC}  & {SROCC} & {RMSE} \\
\hline
\multirow{3}{*}{2D FR IQA} &SSIM  \cite{wang2004ssim} & 0.475	&	0.246	&	1.189 & 0.881& 0.882 & 6.110 & 0.735	&	0.682	&	0.596  \\
\cline{2-11}
&MS-SSIM  \cite{wang2003multiscale} & 0.850	&	0.853	&	0.551 &0.817& 0.788 & 8.946& 0.819	&	0.778	&	0.505\\
\cline{2-11}
&VIF \cite{Sheikh2006}& 0.891	&	0.865	&	0.621 & 0.732& 0.665 & 9.788 &0.816	&	0.784	&	0.508\\
\hline
\multirow{2}{*}{2D NR IQA}& BRISQUE \cite{mittal2012no} & 0.753 &0.814 &0.653 &0.652& 0.642& 13.180 &0.660 &0.614 &0.645\\
\cline{2-11}
&NIQE \cite{mittal2013making}& 0.552	&0.418	&1.032  &0.620& 0.617& 14.113 &0.578	&0.501	&0.718\\
\hline
\hline
\multirow{2}{*}{2D FR VQA}&STMAD \cite{vu2011spatiotemporal} & 0.640 & 0.349 & 0.951& 0.898&0.903 &5.913 & 0.814 & 0.774 & 0.585\\
\cline{2-11}
&VQM \cite{website:vqm}& 0.724&0.702&0.783 &0.905&0.912&5.865 &0.837&0.803	&0.480\\
\hline
\hline
\multirow{2}{*}{3D FR IQA}&STRIQE \cite{khan2015full} & 0.793	&	0.773	&	0.754 & 0.717	&	0.680	&	9.578  & 0.677	&	0.652	&	0.647\\
\cline{2-11}
&MJ3DQA \cite{chen2013full}& 0.798	&	0.786	&	0.746 & 0.787&0.720	&9.484  & 0.725	&	0.686	&	0.605\\
\hline
\hline 
\multirow{1}{*}{3D FR VQA}&$\text{DeMo}_{3D}$ \cite{appina2018}  & 0.927&0.9187&0.456 & 0.903& 0.899& 5.839&0.733 &0.652& 0.698\\
\hline
\multirow{3}{*}{3D NR VQA}&VQUEMODES \cite{appina2018no} & 0.969&0.963&0.263&0.894&0.889&5.912 &0.878&0.839&0.444\\
\cline{2-11}
&$\text{MoDi}_{3D}$ \cite{appina2019subjective} &0.606&0.623&0.985&0.675 & 0.655 & 9.592&0.699&0.623&0.9853\\
\cline{2-11}
&CBSE &0.674&0.653&0.752&0.731&0.6850&9.788&0.743&0.669&0.590\\
\hline
\end{tabular}
\label{table:3DVideoSet}
\end{table*}
\begin{table*}[!htbp]
 \caption{\small Statistical analysis comparison on the performances of off-the-shelf 2D and 3D IQA/VQA models and proposed CBSE model on the VAD stereo video dataset. The symbol `1' indicates the performance of the algorithm in the row is better than the performance of the algorithm in the column, and `0' indicates vice-versa. In each cell, the first two symbols correspond to the fog and haze distortions, and the last symbol represents the overall dataset.}
 \hspace{-0.5cm}
\begin{tabular}{|p {1 cm}| p {0.4 cm}| p {0.75 cm}| p {0.35 cm}| p {0.6 cm}| p {0.4 cm}| p {0.4 cm}| p {0.45 cm}| p {0.45 cm}| p {0.45 cm}| p {0.45 cm}| p {0.45 cm}| p {0.45 cm}| p {0.5 cm}| p {0.77 cm}| p {0.7 cm}| p {0.95 cm}| p {0.6 cm}|p {0.5 cm}|} 
\hline
 \tiny Model  &  \tiny SSIM &\tiny{MS-SSIM}  &\tiny{VIF} &\tiny{BRISQUE}&\tiny{NIQE} &\tiny{PIQE} & \tiny{ILNIQE}&\tiny{STMAD} &\tiny{SpEED} &\tiny{VIIDEO} &\tiny{NSTSS} & \tiny{STRIQE} & \tiny{MJ3DQA} & \tiny{MO-NIQE} &\tiny{$\text{DeMo}_{3D}$} &\tiny VQUEMODES  & \tiny {\tiny{$\text{MoDi}_{3D}$}} & \tiny $\text{CBSE}$\\
\hline
\tiny SSIM &\tiny{- - -}&\tiny{0 0 0}&\tiny{0 1 1}&\tiny{1 1 1}&\tiny{1 1 1}&\tiny{1 1 1}&\tiny{1 1 1}&\tiny{0 0 0}&\tiny{0 0 0}&\tiny{1 1 1}&\tiny{0 1 1}&\tiny{0 0 0}&\tiny{0 0 0}&\tiny{1 1 1}&\tiny{0 0 0}&\tiny{0 0 0}&\tiny{1 1 1}&\tiny{0 0 0}\\
\hline
\tiny{MS-SSIM} &\tiny{1 1 1}&\tiny{- - -}&\tiny{0 1 0}&\tiny{1 1 1}&\tiny{1 1 1}&\tiny{1 1 1}&\tiny{1 1 1}&\tiny{0 0 0}&\tiny{0 0 0}&\tiny{1 1 1}&\tiny{0 1 1}&\tiny{0 0 0}&\tiny{0 0 0}&\tiny{1 1 1}&\tiny{0 0 0}&\tiny{0 0 0}&\tiny{1 1 1}&\tiny{0 0 0}\\
\hline
\tiny{VIF}  &\tiny{1 0 0}&\tiny{1 0 1}&\tiny{- - -}&\tiny{1 1 1 }&\tiny{1 1 1}&\tiny{1 1 1}&\tiny{1 1 1}&\tiny{0 0 0}&\tiny{0 0 0}&\tiny{1 1 1}&\tiny{1 1 1}&\tiny{0 0 0}&\tiny{0 0 0}&\tiny{1 1 1}&\tiny{0 0 0}&\tiny{0 0 0}&\tiny{1 1 1}&\tiny{0 0 0}\\
\hline%
\tiny BRISQUE &\tiny{0 0 0}&\tiny{0 0 0}&\tiny{0 0 0}&\tiny{- - - }&\tiny{0 0 0}&\tiny{0 1 0}&\tiny{0 0 0}&\tiny{0 0 0}&\tiny{0 0 0}&\tiny{0 1 0}&\tiny{0 0 0}&\tiny{0 0 0}&\tiny{0 0 0}&\tiny{0 1 0}&\tiny{0 0 0}&\tiny{0 0 0}&\tiny{0 0 0}&\tiny{0 0 0}\\
\hline
\tiny NIQE &\tiny{0 0 0}&\tiny{0 0 0}&\tiny{0 0 0}&\tiny{1 1 1}&\tiny{- - -}&\tiny{1 1 1}&\tiny{0 1 1}&\tiny{0 0 0}&\tiny{0 0 0}&\tiny{1 1 1}&\tiny{0 1 0}&\tiny{0 0 0}&\tiny{0 0 0}&\tiny{1 1 1}&\tiny{0 0 0}&\tiny{0 0 0}&\tiny{0 0 0}&\tiny{0 0 0}\\
\hline
\tiny PIQE &\tiny{0 0 0}&\tiny{0 0 0}&\tiny{0 0 0}&\tiny{1 0 1}&\tiny{0 0 0}&\tiny{- - -}&\tiny{0 0 0}&\tiny{0 0 0}&\tiny{0 0 0}&\tiny{0 0 0}&\tiny{0 0 0}&\tiny{0 0 0}&\tiny{0 0 0}&\tiny{0 1 1}&\tiny{0 0 0}&\tiny{0 0 0}&\tiny{0 0 0}&\tiny{0 0 0}\\
\hline
\tiny ILNIQE &\tiny{0 0 0}&\tiny{0 0 0}&\tiny{0 0 0}&\tiny{0 0 0}&\tiny{1 0 0}&\tiny{1 1 1}&\tiny{- - -}&\tiny{0 0 0}&\tiny{0 0 0}&\tiny{1 1 1}&\tiny{0 0 0}&\tiny{0 0 0}&\tiny{0 0 0}&\tiny{1 1 1}&\tiny{0 0 0}&\tiny{0 0 0}&\tiny{0 0 0}&\tiny{0 0 0}\\
\hline
\tiny STMAD &\tiny{1 1 1}&\tiny{1 1 1}&\tiny{1 1 1}&\tiny{1 1 1 }&\tiny{1 1 1}&\tiny{1 1 1}&\tiny{1 1 1}&\tiny{- - -}&\tiny{0 0 0}&\tiny{1 1 1}&\tiny{1 1 1}&\tiny{1 1 1}&\tiny{1 0 1}&\tiny{1 1 1}&\tiny{0 0 0}&\tiny{0 0 0}&\tiny{1 1 1}&\tiny{0 0 0}\\
\hline
\tiny SpEED &\tiny{1 1 1}&\tiny{1 1 1}&\tiny{1 1 1}&\tiny{1 1 1 }&\tiny{1 1 1}&\tiny{1 1 1}&\tiny{1 1 1}&\tiny{1 1 1}&\tiny{- - -}&\tiny{1 1 1}&\tiny{1 1 1}&\tiny{1 1 1}&\tiny{1 1 1}&\tiny{1 1 1}&\tiny{1 1 1}&\tiny{0 0 0}&\tiny{1 1 1}&\tiny{1 0 1}\\
\hline
\tiny VIIDEO &\tiny{0 0 0}&\tiny{0 0 0}&\tiny{0 0 0}&\tiny{1 0 1 }&\tiny{1 1 1}&\tiny{1 1 1}&\tiny{0 0 0}&\tiny{0 0 0}&\tiny{0 0 0}&\tiny{- - -}&\tiny{0 0 0}&\tiny{0 0 0}&\tiny{0 0 0}&\tiny{1 1 1}&\tiny{0 0 0}&\tiny{0 0 0}&\tiny{0 0 0}&\tiny{0 0 0}\\
\hline
\tiny NSTSS &\tiny{1 0 0}&\tiny{1 0 0}&\tiny{0 0 0}&\tiny{1 1 1 }&\tiny{1 0 1}&\tiny{1 1 1}&\tiny{1 1 1}&\tiny{0 0 0}&\tiny{0 0 0}&\tiny{1 1 1}&\tiny{- - -}&\tiny{0 0 0}&\tiny{0 0 0}&\tiny{1 1 1}&\tiny{0 0 0}&\tiny{0 0 0}&\tiny{1 0 0}&\tiny{0 0 0}\\
\hline
\tiny STRIQE &\tiny{1 1 1}&\tiny{1 1 1}&\tiny{1 1 1}&\tiny{1 1 1 }&\tiny{1 1 1}&\tiny{1 1 1}&\tiny{1 1 1}&\tiny{0 0 0}&\tiny{0 0 0}&\tiny{1 1 1}&\tiny{1 1 1}&\tiny{- - -}&\tiny{0 0 0}&\tiny{1 1 1}&\tiny{0 0 0}&\tiny{0 0 0}&\tiny{0 0 0}&\tiny{0 0 0}\\
\hline
\tiny MJ3DQA &\tiny{1 1 1}&\tiny{1 1 1}&\tiny{1 1 1}&\tiny{1 1 1}&\tiny{1 1 1}&\tiny{1 1 1}&\tiny{1 1 1}&\tiny{0 1 0}&\tiny{0 0 0}&\tiny{1 1 1}&\tiny{1 1 1}&\tiny{1 1 1}&\tiny{- - -}&\tiny{1 1 1}&\tiny{0 0 0}&\tiny{0 0 0}&\tiny{0 0 0}&\tiny{0 1 0}\\
\hline
\tiny MO-NIQE &\tiny{0 0 0}&\tiny{0 0 0}&\tiny{0 0 0}&\tiny{1 0 1 }&\tiny{0 0 0}&\tiny{1 0 0}&\tiny{0 0 0}&\tiny{0 0 0}&\tiny{0 0 0}&\tiny{0 0 0}&\tiny{0 0 0}&\tiny{0 0 0}&\tiny{0 0 0}&\tiny{- - -}&\tiny{0 0 0}&\tiny{0 0 0}&\tiny{0 0 0}&\tiny{0 0 0}\\
\hline
\tiny{$\text{DeMo}_{3D}$} &\tiny{1 1 1}&\tiny{1 1 1}&\tiny{1 1 1}&\tiny{1 1 1}&\tiny{1 1 1}&\tiny{1 1 1}&\tiny{1 1 1}&\tiny{1 1 1}&\tiny{0 0 0}&\tiny{1 1 1}&\tiny{1 1 1}&\tiny{1 1 1}&\tiny{1 1 1}&\tiny{1 1 1}&\tiny{- - -}&\tiny{0 0 0}&\tiny{1 1 1}&\tiny{0 0 0}\\
\hline
\tiny VQUEMODES &\tiny{1 1 1}&\tiny{1 1 1}&\tiny{1 1 1}&\tiny{1 1 1 }&\tiny{1 1 1}&\tiny{1 1 1}&\tiny{1 1 1}&\tiny{1 1 1}&\tiny{1 1 1}&\tiny{1 1 1}&\tiny{1 1 1}&\tiny{1 1 1}&\tiny{0 0 0}&\tiny{1 1 1}&\tiny{1 1 1}&\tiny{- - -}&\tiny{1 1 1}&\tiny{1 1 1}\\
\hline
{\tiny{$\text{MoDi}_{3D}$}} &\tiny{0 0 0}&\tiny{0 0 0}&\tiny{0 0 0}&\tiny{1 1 1 }&\tiny{1 1 1}&\tiny{1 1 1}&\tiny{1 1 1}&\tiny{0 0 0}&\tiny{0 0 0}&\tiny{1 1 1}&\tiny{0 1 1}&\tiny{1 1 1}&\tiny{0 0 0}&\tiny{1 1 1}&\tiny{0 0 0}&\tiny{0 0 0}&\tiny{- - -}&\tiny{0 0 0}\\
\hline
\tiny {\text{CBSE}} &\tiny{1 1 1}&\tiny{1 1 1}&\tiny{1 1 1}&\tiny{1 1 1 }&\tiny{1 1 1}&\tiny{1 1 1}&\tiny{1 1 1}&\tiny{1 1 1}&\tiny{0 1 0}&\tiny{1 1 1}&\tiny{1 1 1}&\tiny{1 1 1}&\tiny{1 1 1}&\tiny{1 1 1}&\tiny{1 1 1}&\tiny{0 0 0}&\tiny{1 1 1}&\tiny{- - -}\\
\hline 
\end{tabular}
\label{table:statisticalevaluation}
\end{table*}

Figs.~\ref{fig:FogSymmPFeat}, \ref{fig:FogSymmSFeat}, \ref{fig:HazeSymmPFeat} and \ref{fig:HazeSymmSFeat} show the frame wise UGGD feature ($\alpha$, $\beta$) distribution of pristine and its symmetrically distorted versions of fog and haze ambiances of the `RMIT University Courtyard' S3D video sequence computed at first scale, $\theta =  0 ^{\circ}$ and $\Phi =  0 ^{\circ}$ orientation. From the plots, it is clear that ($\alpha$, $\beta$) are able to capture the variations in perceptual quality, and are well segregated with respect to the type and level of distortion. Therefore, these observations have motivated us to use the ($\alpha$, $\beta$) features as perceptual ambiance discriminative features in the proposed CBSE algorithm.
\subsubsection{CBSE score computation}
The proposed CBSE model begins with the generation of Multivariate Gaussian (MVG) models of pristine and distorted videos.

To compute the pristine quality MVG features, we have used the uncompressed video sequences of DML-ITRACK-3D video dataset~\cite{fang2017visual}. We have excluded the uncompressed sequences which we used as reference videos in our subjective study. Therefore, the remaining 36 uncompressed videos from~\cite{fang2017visual} are utilized as a pristine video set in our objective analysis. We first perform the spherical steerable pyramid decomposition on nonoverlapping patches of 36 pristine video sequences at three spatial scales and multiple $\theta$ and $\Phi$ orientations. We then compute the UGGD model fitting coefficients $(\alpha, \beta)$ at each subband of all patches of reference video dataset. The UGGD feature set of pristine S3D videos are represented as follows: 
\begin{align}
L^{P}{(\alpha,\beta)} = [\alpha^{p}_{gh};\beta^{p}_{gh}],
\end{align}
where $g$ represents the number of nonoverlap patches of an S3D video and its maximum value is 144. $h$ indicates the subband level and its range is $(1\leq h \leq 135~(\text{3 scales} \times \text{9}~\theta~\text{orientations} \times \text{5}~\Phi~\text{orientations}))$. $p$ represents the reference video sequence and $P$ is total number of pristine videos. $L^{P}{(\alpha,\beta)}$ represents the pristine quality UGGD feature vector set.

Similar to the pristine uncompressed S3D video set, we have computed the UGGD model fitting coefficients $(\alpha, \beta)$ at the aforementioned scales and orientations of spherical steerable pyramid decomposition of the patches of a distorted S3D video. 
\begin{align}
L^{D}{(\alpha,\beta)} = [\alpha^{D}_{gh};\beta^{D}_{gh}],
\end{align}
where $D$ represents a distorted video and $L^{D}{(\alpha,\beta)}$ indicates the UGGD feature set of $D$.

We have performed MVG fit to the pristine and distorted feature sets, and estimated the mean vector and covariance matrix from each MVG fit. 
\begin{align}
\hspace{-0.3cm}    \mathcal{F}(Y) = \frac{1}{2(\pi)^{\frac{h}{2}} | {\scriptstyle \sum} | ^{\frac{1}{2}}} \times \text{exp}(-\frac{1}{2} (y-\mu)^{T} {{\scriptstyle \sum}}^{-1} (y-\mu)),\\
\Big (\mu^{P}{\big (\alpha,\beta \big)}, \textstyle \sum^{P} \big (\alpha,\beta \big) \Big) = \mathcal{F}(L^{P}{(\alpha,\beta)}),\\
    \Big (\mu^{D}{\big (\alpha,\beta \big)}, \textstyle \sum^{D} \big (\alpha,\beta \big) \Big) = \mathcal{F}(L^{D}{(\alpha,\beta)}),
\end{align}
where $Y$ represents the feature vector and $\mathcal{F}(Y)$ is the MVG density function. ($\mu$, $\scriptstyle \sum$) are mean vector and covariance matrix of $\mathcal{F}(Y)$. $\Big (\mu^{P}{\big (\alpha,\beta \big)},~\scriptstyle \sum^{P} \big (\alpha,\beta \big) \Big)$ and $\Big (\mu^{D}{\big (\alpha,\beta \big)},~\scriptstyle \sum^{D} \big (\alpha,\beta \big) \Big)$ are the estimated MVG model parameters of pristine and distorted feature set. 

We have computed the Bhattacharyya distance measure~\cite{schweppe1967bhattacharyya} between the MVG model parameters of pristine and distorted feature sets to measure the perceptual deviation of a distorted S3D video compared with the pristine S3D video.
\begin{align}
    S_\mu &= \log \left ( \sum_{k=1}^{h} \sqrt{\mu^{P}{\big (\alpha,\beta \big)} \times \mu^{D}{\big (\alpha,\beta \big)}} \right ),\\
    S_{\scriptstyle \sum} &= \log \left ( \sum_{k=1}^{h} \sqrt{\scriptstyle \sum^{P} \big (\alpha,\beta \big) \times \scriptstyle \sum^{D} \big (\alpha,\beta \big)} \right ),
\end{align}
$S_\mu$ and $S_{\textstyle \sum}$ are the Bhattacharyya distance measures computed between mean vectors and covariance matrices of pristine and distorted MVG model parameters. 

The computed distance measures $S_\mu$ and $S_{\scriptstyle \sum}$ increase with the distortion strength. So, we perform the product between both distance measures to estimate the overall quality score of an S3D video. 
\begin{align}
    \text{CBSE}= S_\mu \times S_{\scriptstyle \sum},
\end{align}
where CBSE is the overall quality score of an S3D video. 

Figs.~\ref{fig:PerTheta} and \ref{fig:PerPhi} show the performance of proposed CBSE model in terms of LCC score at each spatial scale and $\theta$ and $\Phi$ orientations. It is evident from the plots that the proposed model demonstrates consistent performance across all scales and orientations. Additionally, it shows significant performance improvements across all subbands when we combine the features of all scales and orientations. 
\section{Results and Discussion}
\label{sec:ResultsandDiscussion}
We evaluate the effectiveness of the proposed objective model on IRCCYN~\cite{Urvoy2012}, LFOVIAPh1~\cite{appina2017subjective}, LFOVIAPh2~\cite{appina2019subjective}, and proposed VAD stereo video datasets. 

The IRCCYN dataset~\cite{Urvoy2012} is a symmetrically distorted S3D video dataset with a combination of 10 pristine and 70 test S3D video sequences. The duration of each video sequence is either 13 sec. or 16 sec. with a frame rate of 25 fps. All video sequences have true HD resolution and are saved in .avi container. The test video sequences are a combination of H.264 compressions and JP2K distortions, and H.264 compression sequences are generated by simulating the JM reference software on the individual left and right videos by changing the quantization parameter. The JP2K distorted videos are developed by varying the bitrate parameter on a frame-by-frame basis of both views of an S3D video. MOS values are published as the final perceptual quality representation of the dataset. 

LFOVIAPh1 and LFOVIAPh2 S3D video datasets~\cite{appina2017subjective,appina2019subjective} are a combination of symmetrically and asymmetrically distorted video sequences. LFOVIAPh1 dataset contains 6 reference and 144 test S3D video sequences, while the LFOVIAPh2 dataset has 12 pristine and 288 test video sequences. In LFOVIAPh1 dataset, the test video sequences are generated by simulating the H.264 compression artefacts by changing the bitrate parameter on the left and right views of an S3D video. Each video sequence has a resolution of $1836 \times 1056$ and a duration of 10 sec. at a frame rate of 25 fps. The LFOVIAPh2 dataset is a combination of H.264 and H.265 compressions, frame freeze and blur distortions. The video sequences have a duration of 10 sec. with a frame rate of 25 fps, and the resolution is $1920 \times 1080$. Both datasets published DMOS values as the final perceptual quality representation of the dataset. The details of the VAD stereo video dataset are explained in Section~\ref{sec:subjectivequalityasessment}.

We compute LCC and SROCC scores and Root Mean Square Error (RMSE) values to measure the performance of the proposed objective model. The LCC score indicates a linear relationship between two variables and SROCC measures the monotonic variation between two variables, while the RMSE value represents the error magnitude between predicted scores and human assessment scores. High LCC and SROCC values indicate a good agreement between human assessment scores and predicted model scores. Lower RMSE represents better accuracy of prediction. The performance statistics are reported after performing a four parameter non-linear logistic fit~\cite{website:LIVE_Database_Report}.

\begin{equation}
f(x)=\frac{z_{1}-z_{2}}{1+\text{exp}({\frac{\zeta-z_{3}}{|z_{4}|}})}+z_{2},
\end{equation}
where $\zeta$ denotes the predicted objective model score, and $z_{1}, z_{2}, z_{3}$ and $z_{4}$ are conditioned to provide a best fit of the predicted objective model scores and human assessment scores.

The performance of the proposed objective model is compared against the performances of off-the-shelf 2D and 3D IQA and VQA algorithms. SSIM~\cite{wang2004ssim}, MS-SSIM~\cite{wang2003multiscale} and VIF~\cite{Sheikh2006} are 2D FR IQA models, and BRISQUE~\cite{mittal2012no}, NIQE~\cite{mittal2013making}, PIQUE~\cite{venkatanath2015blind}, and ILNIQE~\cite{zhang2015feature} are 2D NR IQA models. The performances of these algorithms are computed on each frame of left and right videos, and a mean score of all frame-level predictions of both views is computed to estimate the overall quality score of an S3D video. STMAD~\cite{vu2011spatiotemporal}, SpEED~\cite{7979533}, VIIDEO~\cite{mittal2015completely} and NSTSS~\cite{9059006} are 2D FR and NR VQA models. The performances of these algorithms are computed on both left and right views, and an average score of both view scores is estimated to measure the overall S3D video quality. STRIQE~\cite{khan2015full}, MJ3DQA~\cite{chen2013full} and MO-NIQE~\cite{appina2020complete} are 3D FR and NR IQA models. The performance of these algorithms is computed on each frame of an S3D video, and a mean score of all frame-level predictions is computed to estimate the final quality score. $\text{DeMo}_\text{3D}$~\cite{appina2018}, VQUEMODES~\cite{appina2018no} and $\text{MoDi}_\text{3D}$~\cite{appina2019subjective} are S3D FR and NR VQA models. 
\begin{figure*}[!htbp]%
\captionsetup[subfigure]{justification=centering}
\centering
\begin{subfigure}[b]{0.245\textwidth}
\includegraphics[height=2.5cm,width=4cm]{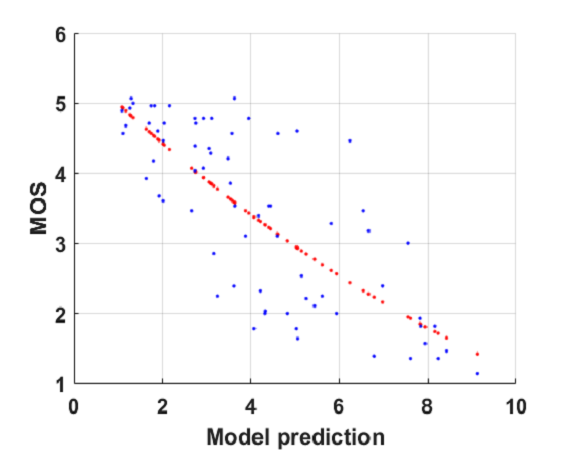}
\subcaption{\small IRCCYN.}
\label{fig:DistScatterIRCCYN} 
\end{subfigure}
\begin{subfigure}[b]{0.245\textwidth}
\includegraphics[height=2.5cm,width=4cm]{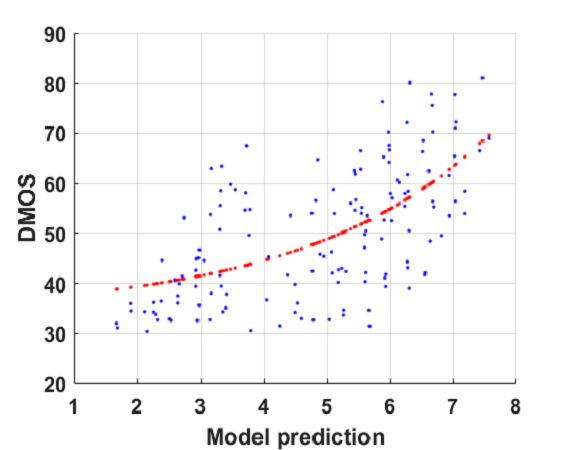}
\subcaption{\small LFOVIAPh1.} 
\label{fig:DistScatterLPh1}
\end{subfigure}
\begin{subfigure}[b]{0.245\textwidth}
\includegraphics[height=2.5cm,width=4cm]{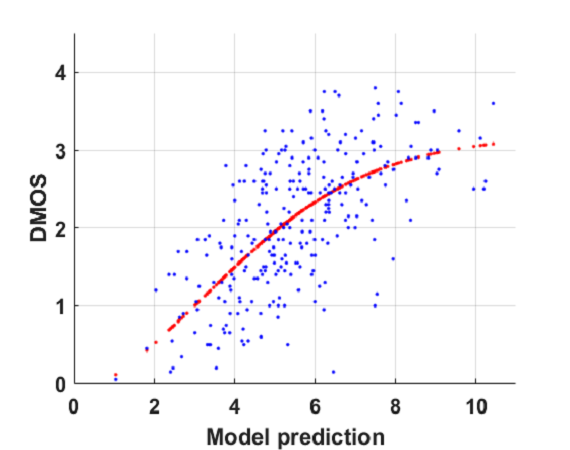}
\subcaption{\small LFOVIAPh2.} 
\label{fig:DistScatterLPh2}
\end{subfigure}
\begin{subfigure}[b]{0.245\textwidth}
\includegraphics[height=2.5cm,width=4cm]{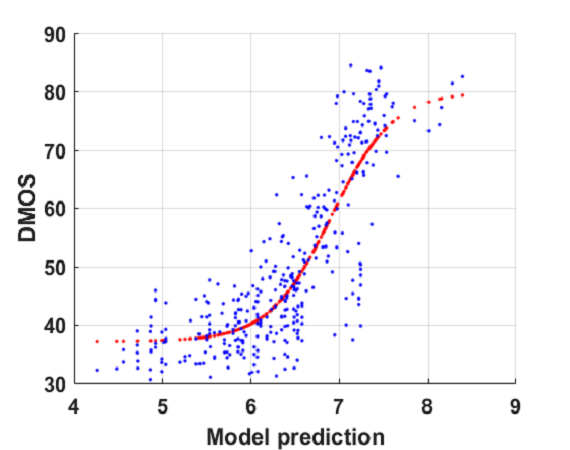}
\subcaption{\small VAD Stereo.} 
\label{fig:DistScatterVAD}
\end{subfigure}
\caption{Scatter plots of proposed algorithm scores versus human assessment scores of the IRCCYN, LFOVIAPh1, LFOVIAPh2 and VAD stereo S3D video datasets.}
\label{fig:SetScatter}
\end{figure*}

Tables~\ref{table:VADStereo} and \ref{table:VADStereoEach} show the performance evaluation of CBSE model on the proposed VAD stereo video dataset. From the tables, it is clear that the proposed algorithm delivers consistent performance across all distortion types and also on symmetrically and asymmetrically distorted videos of the dataset. Further, the proposed algorithm demonstrates competitive performance against 2D and 3D FR and supervised NR IQA and VQA models such as SSIM \cite{wang2004ssim}, MS-SSIM~\cite{wang2003multiscale}, VIF~\cite{Sheikh2006}, BRISQUE~\cite{mittal2012no}, STMAD~\cite{vu2011spatiotemporal}, SpEED~\cite{7979533}, STRIQE~\cite{khan2015full}, MJ3DQA~\cite{chen2013full}, $\text{DeMo}_\text{3D}$~\cite{appina2018} and VQUEMODES~\cite{appina2018no} models. Additionally, it offers superior performance compared to the conventional 2D and 3D unsupervised NR IQA and VQA (NIQE~\cite{mittal2013making}, PIQUE~\cite{venkatanath2015blind}, ILNIQE~\cite{zhang2015feature}, MO-NIQE~\cite{appina2020complete} and $\text{MoDi}_\text{3D}$~\cite{appina2019subjective}) algorithms.

Table~\ref{table:3DVideoSet} shows the performance results of the proposed algorithm on the IRCCYN, LFOVIAPh1 and LFOVIAPh2 S3D video datasets. We evaluate the proposed algorithm on all types of distortions and on symmetrically and asymmetrically distorted versions. From the results, it is clear that the proposed model delivers robust and consistent performance across all distortion types of S3D video datasets. Also, it shows competitive performance against 2D and 3D FR and supervised IQA and VQA models, and demonstrates state-of-the-art performance against 2D and 3D unsupervised NR IQA and VQA algorithms. 

In meteorological terms, fog distortion refers to the formation of a visually whitish/grayish opaque murkiness by the suspension of water droplets in the atmosphere. Haze is a phenomenon that occurs due to the suspension of dry particles, dust, smoke, etc. in the air which yields a yellowish/bluish scene. The intensive colour alterations in the translucent environment of a hazy scene produces severe perceptual depth degradations due to binocular rivalry, but similar deviations are not reported in fog S3D videos. Due to this, 3D IQA and VQA models reported slightly diminished performance on haze S3D videos compared to the fog videos.

Table~\ref{table:statisticalevaluation} shows the comparison of statistical analysis results of the objective model scores to determine whether the LCC scores are significantly different from one another. We followed the method suggested by Moorthy~\textit{et al.}~\cite{moorthy2012video} and Sheikh~\textit{et al.}~\cite{sheikh2006statistical} to perform this analysis, and computed an F-statistic score between the residuals that were produced by two objective quality assessment algorithms after performing the non-linear logistic fit~\cite{website:LIVE_Database_Report}. In the table, if a symbol appears as `1’, it represents that the algorithm indicated in the row performs significantly better than the algorithm indicated in the column, and the opposite occurs when the symbol appears as `0'. The results clearly demonstrate that the proposed algorithm shows competitive performance compared to the state-of-the-art objective algorithms. 

Fig.~\ref{fig:SetScatter} shows scatter plots of the CBSE model predicted scores versus human assessment scores of IRCCYN, LFOVIAPh1, LFOVIAPh2 and VAD stereo video datasets. It is clear that the proposed model correlates well with the human assessment scores, and these plots support the effectiveness of the proposed model. 
\vspace{-0.4cm}
\section{Conclusion}
\label{sec: conclusion}
We proposed subjective and objective methods for S3D videos in this article. In the subjective study, we have created an S3D video dataset with a combination of symmetrically and asymmetrically distorted versions of fog and haze ambiance videos. The dataset is composed of 12 pristine S3D videos and 360 test stimuli. Further, we involved 24 subjects to perform a comprehensive subjective evaluation, and the assessment was done based on ACR-HR protocol. In objective evaluation, we proposed an `OU-DU’ (i.e. completely blind) NR VQA model for S3D videos. We first generated cyclopean frames from the left and right video frames of an S3D video and studied the NSS characteristics at the patch level of pristine and distorted videos. We evaluated this analysis at multiple spatial scales and multiple orientations of the spherical steerable pyramid decomposition. Further, we performed UGGD modeling on subband decompositions and showed the efficacy of UGGD model parameters in discriminating the perceptual quality of an S3D video. The performance of the proposed CBSE model was evaluated on the popular S3D video datasets, and it delivered consistent performance across all distortion types of the datasets. Also, it showed competitive performance against off-the-shelf 2D and 3D IQA and VQA models, even though the proposed model does not perform any training and testing sessions on the video features and the corresponding quality representations. We plan to make the dataset, human assessment scores, and objective method accessible to the research community. In the near future, we will extend the proposed CBSE model to Virtual Reality and Augmented Reality videos. 
\vspace{-0.4cm}

\bibliographystyle{ieeetr}
\footnotesize
\bibliography{mybib}
\end{document}